\documentclass[fleqn,usenatbib]{mnras}
\usepackage{newtxtext,newtxmath}
\usepackage[T1]{fontenc}
\usepackage{ae,aecompl}
\usepackage{graphicx}	
\usepackage{amsmath}	
\usepackage{amssymb}	
\usepackage{soul}
\usepackage{newtxtext}
\usepackage[english]{babel}
\usepackage{booktabs}
\usepackage{graphicx,color}
\usepackage{epstopdf}

\newcommand{\mapsim}{\textsc{MapSim} }
\newcommand{\lcdm}{$\mathrm{\Lambda CDM}$ }
\newcommand{\dust}{{\small DUSTGRAIN} }
\newcommand{\dustp}{{\small DUSTGRAIN}-\emph{pathfinder} }
\newcommand{\dustpp}{{\small DUSTGRAIN}-\emph{pathfinder}}

\newcommand\reallywidehat[1]{\arraycolsep=0pt\relax
\newcommand{\stkout}[1]{\ifmmode\text{\sout{\ensuremath{#1}}}\else\sout{#1}\fi}
\begin{array}{c}
\stretchto{
  \scaleto{
    \scalerel*[\widthof{\ensuremath{#1}}]{\kern-.5pt\bigwedge\kern-.5pt}
    {\rule[-\textheight/2]{1ex}{\textheight}} 
  }{\textheight} %
}{0.5ex}\\           
#1\\                 
\rule{-1ex}{0ex}
\end{array}
}
\title[PLC: MG and Massive Neutrinos]{Weak Lensing Light-Cones in Modified Gravity simulations with and without Massive Neutrinos}
\author[Giocoli C. et al. 2018]{\parbox{\textwidth}{Carlo
    Giocoli$^{1,2,3}$\thanks{E-mail:\href{carlo.giocoli@unibo.it}
      {carlo.giocoli@unibo.it}}, Marco Baldi$^{1,2,3}$,
    Lauro Moscardini$^{1,2,3}$}\\ \\
  $^1$Dipartimento di Fisica e Astronomia, Alma Mater Studiorum
  Universit\`{a} di Bologna, via Gobetti 93/2, I-40129 Bologna, Italy\\
  $^2$INAF - Osservatorio di Astrofisica e Scienza dello Spazio di Bologna,  via Gobetti
  93/3, I-40129 Bologna, Italy \\
  $^3$INFN - Sezione di Bologna, viale Berti Pichat 6/2, I-40127
  Bologna, Italy}

\pubyear{2018}

\begin{document}
\label{firstpage}
\pagerange{\pageref{firstpage}--\pageref{lastpage}}
\maketitle

\begin{abstract}

We present a novel suite of cosmological N-body simulations called the
{\dustpp}, implementing simultaneously the  effects of an extension to
General  Relativity   in  the  form   of  $f(R)$  gravity  and   of  a
non-negligible  fraction  of  massive   neutrinos.   We  describe  the
generation of  simulated weak  lensing and cluster  counts observables
within  a  past  light-cone  extracted from  these  simulations.   The
simulations  have been  performed by  means  of a  combination of  the
{\small MG-GADGET} code and a particle-based implementation of massive
neutrinos, while the light-cones have been generated using the \mapsim
pipeline  allowing  us   to  compute  weak  lensing   maps  through  a
ray-tracing  algorithm  for  different  values  of  the  source  plane
redshift.  The mock observables extracted from our simulations will be
employed for a series of papers focussed on understanding and possibly
breaking  the  well-known   observational  degeneracy  between  $f(R)$
gravity  and massive  neutrinos,  i.e.  the  fact  that some  specific
combinations of the characteristic  parameters for these two phenomena
(the $f_{R0}$  scalar amplitude  and the  total neutrino  mass $\Sigma
m_{\nu}$)  may  result  indistinguishable   from  the  standard  \lcdm
cosmology   through  several   standard   observational  probes.    In
particular,  in the  present  work  we show  how  a {\em  tomographic}
approach to weak lensing statistics  could allow -- especially for the
next generation  of wide-field surveys  -- to disentangle some  of the
models  that appear  statistically indistinguishable  through standard
single-redshift weak lensing probe.
  
\end{abstract}


\section{Introduction}
\label{intro}

General Relativity  predicts that  light-rays are  bent by  the matter
density  distribution  analogously  to  what happens  in  optics  when
light-rays travel through media characterised by different diffraction
indices \citep{einstein18,landau71}.

Light   emitted  by   a  distant   galaxy,  traveling   for  different
gigaparsecs, tends  to be  deflected multiple  times depending  on the
intervening matter density distribution  along the line-of-sight. This
effect  --  termed gravitational  lensing  --  can  be observed  as  a
modification of the intrinsic shape of the background galaxy caused by
the different paths followed by  the various light-rays emitted by its
various components \citep{bartelmann01,bartelmann10}.

When  rays pass  close to  a very  massive and  compact object  -- the
centre of  a massive galaxy or  a galaxy cluster --  they are strongly
deflected  and sometimes  even  follow multiple  paths. The  resulting
galaxy  image appears  highly distorted  into a  gravitational arc  or
multiply  imaged  in different  positions,  depending  on the  optical
lensing  configuration: relative  position between  the observer,  the
lens  and the  source \citep{broadhurst05b,coe13,meneghetti13}.   This
regime  is termed  strong gravitational  lensing.  When  light bundles
travel sufficiently far  from any compact mass  distribution they tend
to be only  slightly deflected \citep{bartelmann10}. In  this case the
intrinsic shape of  the source is only marginally modified:  we are in
the  regime  of  weak  gravitational  lensing.   Considering  that  by
definition the weak lensing effect is tiny, it is necessary to average
on a large number of background sources in order to quantify it and to
indirectly  perform a  reasonable  reconstruction  of the  intervening
matter  density  distribution  along the  line-of-sight  \citep[for  a
  review  see][]{kilbinger14}.  In  this  direction  many ongoing  and
planned  future surveys  are  aimed to  collect  the largest  possible
number   of  galaxy   images  to   measure  their   ellipticities  and
statistically  infer  their tiny  distortions  so  to obtain  a  solid
estimate  of   the  weak  lensing  effect.    The  future  ESA-mission
Euclid\footnote{\href{http://www.euclid-ec.org}{http://www.euclid-ec.org}}
\citep{euclidredbook}  is  expected to  increase  the  number of  weak
lensing sources  by more than a  factor of four/five and  cover a much
larger area  with respect to  ground-based wide-field surveys  -- like
CFHTLenS  \citep{erben12,benjamin13} and  KiDS \citep{hildebrandt17}--
and  to extend  their distribution  to significantly  higher redshifts
\citep{kitching16}.   All  this  will translate  in  an  unprecedented
capability   of  performing   systematic  analyses   of  cosmic   weak
gravitational lensing,  on large  field-of-view, and  accurately trace
the growth of  non-linear structures down to small  scales.  The large
number of background galaxies will  allow also to perform weak lensing
tomographic analyses, on a large  number of redshift bins, tracing the
growth of structures up to high redshifts.

Gravitational lensing  probes the dynamics  and the kinematics  of the
Universe  and  its large-scale  structures,  and  for this  reason  it
represents  an   unbiased  tool  to  probe   gravity  on  cosmological
distances. The need  to test General Relativity  (hereafter GR) beyond
the small  scales of the  Solar System  has recently become  an urgent
task  for the  scientific community,  mainly with  the aim  of finding
possible alternative scenarios to  the simple cosmological constant as
a  source of  the  late  time accelerated  expansion  of the  Universe
\citep{novikov16b,Hu_Sawicki_2007,bertschinger08,schmidt08}.   In this
context,  in order  to be  viable a  Modified Gravity  (MG, hereafter)
model must produce an expansion history that does not deviate too much
from  that  of  the  standard  \lcdm  cosmology,  while  allowing  for
deviations in the dynamical evolution of density perturbations through
gravitational instability. Such deviations  need nonetheless to vanish
(and  GR must  be recovered)  in our  local neighborhood  in order  to
fulfill  the  tight bounds  on  GR  derived  within the  Solar  System
\citep[][]{Bertotti_Iess_Tortora_2003,Will_2005}.   The  mechanism  to
achieve such recovery is generally known as {\em Screening} \citep[see
  e.g.][]{Deffayet_etal_2002,Khoury_Weltman_2004,Hinterbichler_Khoury_2010,Brax_Valageas_2014}
and acts at non-linear scales as a suppression of the relative size of
the MG fifth-force when the  gravitational potential or its derivative
become large.   In this  context, numerical  simulations of  MG models
\citep[see
  e.g.][]{Oyaizu_etal_2008,Schmidt_etal_2009,Ecosmog,Puchwein_Baldi_Springel_2013,Llinares_Mota_Winther_2014,Arnold_etal_2018}
represent the  best phenomenological  laboratories to study  'where to
look at' to find deviations from  GR, as they properly account for the
full non-linear  evolution of structures  that is responsible  for the
onset  of  the {\em  Screening}  effect.   In particular,  statistical
analyses  of gravitational  lensing within  simulated light-cones  are
able to  probe, as  a function  of the  angular scale,  the integrated
effect of  the fifth force caused  by a given modification  of gravity
from  linear  down  to  non-linear  scales  \citep{harnois-deraps15c}.
Recently, \citet{barreira17} have constructed past-light-cones from MG
simulations  finding that  non-standard physics  can leave  signatures
both in  the cosmic shear power  spectrum and in the  projected galaxy
density  profile.   Other   interesting  results  on  past-light-cones
simulations   of   f(R)   models    have   been   also   obtained   by
\citet{higuchi16,shirasaki17,li18}.

The recent tensions between high- and low-redshift probes advocated by
the  Planck   CMB  \citep{planck1_11,planck1_14,planck1_15,planck16a},
Planck  cluster counts  \citep{planckxi,planckxx,planckxxiv} and  weak
lensing               surveys              like               CFHTLens
\citep{fu08,heymans13,kilbinger13,kitching14}         and         KiDS
\citep{hildebrandt17} have  increased the interest  about non-standard
models  as possible  solution  for attenuating  those tensions.   Many
works  \citep{lesgourgues06,costanzi14,poulin18}  have suggested  that
the   presence  of   massive   neutrinos  can   help  reducing   these
tensions. Solar neutrino oscillations \citep{Cleveland_etal_1998} have
revealed the presence of massive neutrinos families, suggesting that a
(yet  undetermined)  fraction  of  the total  matter  density  of  the
Universe must be associated with the cosmic neutrino background.

In recent years,  significant progress has been made  in including the
effects of  massive neutrinos into cosmological  N-body codes employed
to study structure  formation processes from the linear  to the highly
non-linear  regime  in  the  context  of  standard  \lcdm  cosmologies
\citep{Viel_Haehnelt_Springel_2010,Wagner_Verde_Jimenez_2012,Zennaro_etal_2017,Villaescusa-Navarro_etal_2018}.
In  this  work  we  extend  such analysis  to  MG  cosmologies  for  a
particular  choice  of  a  MG  scenario,  namely  the  $f(R)$  gravity
\citep[][]{Buchdahl_1970,Starobinsky_1980,Hu_Sawicki_2007,Sotiriou_Faraoni_2010}
by  developing  a  series  of N-body  simulations  performed  with  an
extended   version   of   the    modified   gravity   code   MG-GADGET
\citep{Puchwein_Baldi_Springel_2013} which  includes at the  same time
the particle-based  implementation of  massive neutrinos  developed by
\citet{Viel_Haehnelt_Springel_2010}.      Some     early     numerical
investigations     performed      with     such      extended     code
\citep[see][]{Baldi_etal_2014} have highlighted a strong observational
degeneracy between the effects of  $f(R)$ gravity and those of massive
neutrinos on structure formation processes over a wide range of scales
and redshifts,  covering both  the linear  and non-linear  regimes. In
this  context, a  degeneracy means  the property  of two  cosmological
models  characterised by  fundamentally different  laws of  physics or
energy  content   being  indistinguishable  from  each   other  within
observational errors.   In particular,  \citet{Baldi_etal_2014} showed
that  a proper  combination of  $f(R)$ gravity  parameters and  of the
total neutrino mass $\Sigma m_{\nu  }$ may result in basic observables
like the non-linear matter power spectrum, the halo abundance, and the
halo bias to be statistically  consistent with \lcdm.  Testing the use
of  weak  lensing tomography  as  a  probe to  disentangle  degenerate
cosmological scenarios  will be one of  the main goals of  the present
work.   Our simulations  --  called the  \dustp --  will  allow us  to
investigate the joint effects of $f(R)$ gravity, and massive neutrinos
in  the non-linear  regime of  structure formation  and systematically
study particular  non-standard degenerate cosmological  models through
weak lensing and halo counts statistics  with the aim of shedding some
light onto  which statistics may  help us disentangle  such degenerate
models from the standard \lcdm one.  We based our investigation on the
analysis  of past-light-cones  extracted from  the simulations  within
which we  computed the galaxy  cluster distribution and  the projected
cosmic shear maps.  The simulations presented here will be the base of
a series of papers focussing on various different observational probes
\citep[][]{peel18,hagstotz18}.

Our  paper  is  organized  as follows.   In  Section~\ref{simMain}  we
describe our numerical  simulations, their post-processing procedures,
and    the    past-light-cones     constructed    in    the    various
runs. Section~\ref{results} is  devoted to discuss our  results on the
halo  mass functions  in the  past-light-cone (\ref{hmf})  and on  the
properties of the convergence maps (\ref{convergence}): power spectra,
tomographic analyses and one point statistics.  In Section~\ref{final}
we summarise our results and draw our conclusions.

\section{Numerical Simulations}
\label{simMain}

\subsection{The \dustp simulations}
\label{sim}

For the  analysis discussed  in the  present work we  make use  of the
projected  matter distribution  and halo  catalogues extracted  from a
suite of  cosmological Dark Matter-only simulations  called the \dustp
runs.  The  \dust (Dark  Universe Simulations to  Test GRAvity  In the
presence  of Neutrinos)  project  is an  ongoing  enterprise aimed  at
producing large  and detailed mock observations  of galaxy clustering,
weak  lensing,  CMB  lensing  and redshift-space  distortions  in  the
context of cosmological models characterised  by a modification of the
laws of gravity  from their standard General Relativity form  and by a
non-negligible fraction  of the  cosmic matter  density being  made of
standard  massive neutrinos.   For the  former we  consider an  $f(R)$
gravity theory defined by the action \citep[][]{Buchdahl_1970}:
\begin{equation}
\label{fRaction}
  S = \int {\rm d}^4x \, \sqrt{-g} \left( \frac{R+f(R)}{16 \pi G} +
  {\cal L}_m \right),
\end{equation}
where we  assume for  the $f(R)$ function  the widely  considered form
\citep[][]{Hu_Sawicki_2007}:
\begin{equation}
\label{fRHS}
f(R) = -m^2 \frac{c_1 \left(\frac{R}{m^2}\right)^n}{c_2
  \left(\frac{R}{m^2}\right)^n + 1},
\end{equation}
with $R$ being  the Ricci scalar curvature, $ m^2  \equiv H_0^2 \Omega
_{\rm M}$  being a mass scale,  while $c_{1}$, $c_{2}$, {and  $n$} are
non-negative constant free  parameters of the model. Here  we focus on
the specific case for which  the background expansion history is fixed
to the standard \lcdm one by choosing $c_{1}/c_{2} = 6\Omega _{\Lambda
}/\Omega _{\rm  M}$ --  where $\Omega_{\Lambda}$ and  $\Omega_{\rm M}$
represent the vacuum and matter energy density respectively, under the
condition  $c_2 (R/m^2)^n  \gg 1$,  so that  the scalar  field $f_{R}$
takes the approximate form:
\begin{equation}
  f_R \approx -n \frac{c_1}{c_2^2}\left(\frac{m^2}{R}\right)^{n+1}.
\label{eq:fR-R,n_relation}
\end{equation}

We restrict our analysis  to the case $n=1$ so that  the model is left
with only one free parameter, which can be expressed as:
\begin{equation}
f_{R0}\equiv -\frac{1}{c_{2}}\frac{6\Omega _{\Lambda }}{\Omega
  _{M}}\left( \frac{m^{2}}{R_{0}}\right) ^{2}.
\end{equation}

As         it         is        now         generally         accepted
\citep[][]{Baldi_etal_2014,He_2013,Motohashi_etal_2013,Wright_Winther_Koyama_2017}
Modified  Gravity  theories  such  as  e.g.   $f(R)$  gravity  in  the
\citeauthor{Hu_Sawicki_2007} form  are strongly  degenerate in  a wide
range  of their  observable  footprints with  the  effects of  massive
neutrinos on structure formation, posing serious challenges to present
and  future  large galaxy  surveys  in  devising robust  and  reliable
methods  to disentangle  the two  phenomena.  In  particular, standard
statistics  such  as  the  matter  auto-power  spectrum,  the  lensing
convergence power spectrum,  and the halo mass function  may be hardly
distinguishable  from  their  standard   \lcdm  expectation  for  some
specific combinations of the $f(R)$  gravity parameter $f_{R0}$ and of
the total neutrino mass $m_{\nu  }\equiv \Sigma m_{\nu ,i}$ \citep[see
][]{Baldi_etal_2014,peel18}  which  define  a  ``maximum  degeneracy''
relation between the two models.

As the degeneracy is mostly driven by the non-linear behaviour of both
the Modified  Gravity and the  massive neutrinos effects  on structure
formation,  linear tools  are  not suitable  to  properly explore  the
combined  parameter space  and identify  the shape  of such  ``maximum
degeneracy''  relation.   Therefore,  in  order to  set  the  specific
parameter combinations for the full \dust runs it was necessary to run
a suite  of smaller-scale  and lower-resolution N-body  simulations --
called the {\em pathfinder} runs --  to (at least coarsely) sample the
$\left\{ f(R), m_{\nu}\right\}$ parameter  space. While the full-scale
\dust simulations are still running at the time of writing this paper,
the  {\em  pathfinder}  runs  already provide  a  wealth  of  valuable
simulated data and of novel  information content to deserve a detailed
analysis. We start such investigations with the present work focussing
on weak lensing tomography on the past light-cone and its correlations
with  the distribution  of  massive clusters  and  with two  companion
papers focussing on higher-order statistics of weak lensing and on the
modelling   of   the  halo   mass   function   in  these   cosmologies
\citep[][respectively]{hagstotz18,peel18}. These will be followed soon
by a series of other papers targeted at other types of observables.

\begin{figure*}
\includegraphics[width=0.55\hsize]{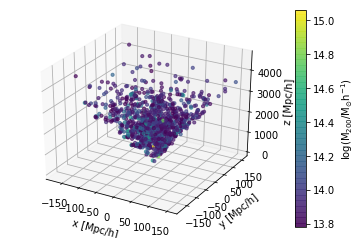}
\includegraphics[width=0.44\hsize]{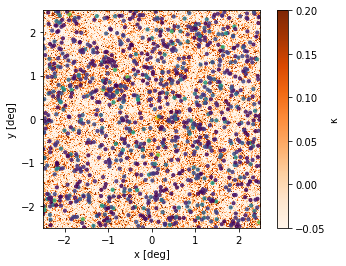}  
\caption{Schematic  representation of  the  past-light-cone using  our
  \mapsim routine. Left panel shows the three-dimensional distribution
  of haloes within the light-cone with 5$\times$5 sq. deg. aperture up
  to  redshift $z=4$.   We display  all haloes  with mass  larger than
  $M_{200}\geq 6\times 10^{13} M_{\odot}/h$  colour coded according to
  their mass.   Right panel shows  the convergence map for  $z_s=4$ --
  which represents the  base of our past-light-pyramid --  on which we
  display also the projected distribution of the haloes present inside
  the field-of-view. The scale color  of the underlining map refers to
  the convergence value corresponding to each pixel.
  \label{FigSk}}
\end{figure*}

In the {\em Newtonian} gauge, we can write the perturbed metric as:
\begin{equation}
\mathrm{d}s^2 =  (1+2 \Psi) \mathrm{d}t^2 - a^2(t) ( 1 - 2 \Phi)  
\mathrm{d}x_i \mathrm{d}x_j\,,
\end{equation}
where $\Phi$  and $\Psi$  are the Einstein-frame  metric gravitational
potentials. In standard  GR model, where the  accelerated expansion of
the Universe is driven by a  cosmological constant we can read the two
potentials as:
\begin{equation}
\Phi = \Psi = \Psi_N\,,
\end{equation}
where the  Newton potential $\Psi_N$  can be  written in terms  of the
Poisson equation as:
\begin{equation}
\frac{\nabla^2 \Psi_N}{a^2}  = 4 \pi G \delta \rho = \frac{3 \Omega_m H_0^2}{2 a^3} \delta .
\end{equation}
The symbols  $\Omega_M$ and $H_0$  represent the total  matter density
parameter and the  Hubble constant at the  present time, respectively,
and  $\delta\equiv\delta  \rho /  \bar{\rho}$  is  the matter  density
contrast.\\ In the $f(R)$ modified gravity models the variation of the
Einstein equations leads to the two modified Poisson equations for the
metric potentials:
\begin{equation}
\frac{\nabla^2 \Phi}{a^2} = - \frac{c^2 \nabla^2}{2 a^2} \delta f_R + 4 \pi G \delta \rho\,,
\end{equation}
\begin{equation}
\frac{\nabla^2 \Psi}{a^2} =  \frac{c^2 \nabla^2}{2 a^2} \delta f_R + 4 \pi G \delta \rho\,,
\end{equation}
which can be written in terms of the Newton potential as:
\begin{equation}
\Phi = \Psi_N - \frac{c^2}{2} \delta f_R\,,\;\;\;
\Psi = \Psi_N + \frac{c^2}{2} \delta f_R\,.
\end{equation}
At this point we can introduce the lensing potential $\Phi_{l}$ as:
\begin{equation}
\Phi_{l} = \frac{\Phi + \Psi}{2}\,
\end{equation}
from which we can notice that the  two extra terms in MG models cancel
out: the  lensing potential in  $f(R)$ gravity remains  unchanged from
its  standard  GR  form.  Therefore, we  can  define  the  convergence
$\kappa$ in the usual way as:
\begin{equation}
\kappa (\theta) = \int_0^{\infty} \mathrm{d}w \frac{w}{c^2} 
g(w) \nabla^2 \Phi_l(w,w \theta)\,,
\end{equation}
where  $w$ represents  the  comoving radial  distance  and $g(w)$  the
survey weight function.

Coming to a technical description  of the \dust-{\em pathfinder} runs,
these  are   cosmological  collisionless  simulations   following  the
evolution of  an ensemble of  $768^{3}$ Dark Matter particles  of mass
$m_{\rm  CDM}= 8.1\times  10^{10}$  M$_{\odot }/h$  (for  the case  of
$m_{\nu  }=0$) and  of as  many neutrino  particles (for  the case  of
$m_{\nu }>0$) within a periodic  cosmological box of $750$ Mpc$/h$ per
side,  under the  effect of  a gravitational  interaction dictated  by
eq.~(\ref{fRaction}). Standard  cosmological parameters are set  to be
consistent      with      the      Planck      $2015$      constraints
\citep[][]{Planck_2015_XIII},  namely  $\Omega _{\rm  M}=\Omega  _{\rm
  CDM}+\Omega  _{\rm b}+\Omega  _{\nu} =  0.31345$, $\Omega  _{\rm b}=
0.0481$,  $\Omega _{\Lambda  }= 0.68655$,  $H_{0}= 67.31$  km s$^{-1}$
Mpc$^{-1}$, ${\cal{A}} _{\rm  s}= 2.199\times 10^{-9}$, $n_{s}=0.9658$
which  give  at  $z=0$  a   root-mean-square  of  the  linear  density
fluctuation   smoothed  on   a   scale  of   $8$   Mpc$/h$  equal   to
$\sigma_8=0.847$.

For  these  simulations  we   employed  the  {\small  MG-Gadget}  code
\citep[see][]{Puchwein_Baldi_Springel_2013}, a modified version of the
      {\small  GADGET} code  \citep[][]{gadget-2} that  implements the
      extra  force   and  the  {\em  Chameleon}   screening  mechanism
      \citep[see][]{Khoury_Weltman_2004}   that  characterize   $f(R)$
      gravity by solving the  non-linear Poisson-like equation for the
      $f_{R}$ scalar degree of freedom
\begin{equation}
\label{modPoisson}
  \nabla^2 f_R = \frac{1}{3}\left(\delta R - 8 \pi G \delta \rho
  \right) \,,
\end{equation}
through  a Newton-Gauss-Seidel  (NGS) iterative  scheme on  the native
gravitational  Tree  of  {\small  GADGET} which  is  exploited  as  an
adaptive   mesh    \citep[see][for   more   details    about   {\small
    MG-GADGET}]{Puchwein_Baldi_Springel_2013}.    {\small   MG-Gadget}
code has been extensively tested  \citep[see e.g. the Modified Gravity
  code  comparison  project  described in  ][]{Winther_etal_2015}  and
employed in the recent past for a wide variety of applications ranging
from     large-scale     collisionless    cosmological     simulations
\citep[][]{Baldi_Villaescusa-Navarro_2018,Arnold_etal_2018}         to
hydrodynamical                                             simulations
\citep[][]{Arnold_Puchwein_Springel_2014,Arnold_Puchwein_Springel_2015,Roncarelli_Baldi_Villaescusa-Navarro_2018},
to     zoomed    simulations     of     Milky    Way-sized     objects
\citep[][]{Arnold_Springel_Puchwein_2016,Naik_etal_2018}.    In   this
work,     following     the     approach    already     adopted     in
\citet{Baldi_etal_2014} we have combined the {\small MG-Gadget} solver
with the particle-based implementation  of massive neutrinos developed
by  \citet{Viel_Haehnelt_Springel_2010} in  order  to include  massive
neutrinos in our simulations as an additional family of particles with
its  specific initial  transfer  function  and velocity  distribution.
Therefore, both  $\mathrm{CDM}$ and  neutrino particles  contribute to
the density source term for  the scalar perturbations evolution on the
right-hand side of eq.~(\ref{modPoisson}).

Initial conditions have been generated  following the approach of e.g.
\citet{Zennaro_etal_2017,Villaescusa-Navarro_etal_2018}  which amounts
to generating two  fully correlated random realisations  of the linear
matter  power spectrum  for standard  Cold Dark  Matter particles  and
massive  neutrinos  based  on   their  individual  transfer  functions
computed by  the linear Boltzmann code  {\small CAMB} \citep[][]{camb}
at  the  starting  redshift  of  the  simulation  $z_{i}=99$,  and  by
computing  the scale-dependent  growth rate  $D_{+}(z_{i},k)$ for  the
neutrino   component   in   order  to   correctly   compute   neutrino
gravitational velocities.  Thermal neutrino  velocities are then added
on  top  of  the  latter  by random  sampling  the  neutrino  momentum
distribution  at  the  starting  redshift  $z_{i}$  for  the  specific
neutrino mass under consideration.

The \dustp  simulations spanned the parameter  range $-1\times 10^{-4}
\le f_{R0} \le -1\times 10^{-6}$  for the scalar amplitude and $0~{\rm
  eV} \le m_{\nu  } \le 0.3~{\rm eV}$ for the  neutrino mass with $20$
different  parameter  combinations.  In  this  work,  we restrict  the
analysis to  a subset of  these simulations including $3$  pure $f(R)$
runs (i.e. for $m_{\nu } = 0$  eV) and those of their massive neutrino
counterparts  that result  in a  significant observational  degeneracy
amounting to one or two values  of $m_{\nu }$ for each $f_{R0}$ value,
plus a standard \lcdm simulation (i.e. GR and massless neutrinos) as a
reference model, for a total of $9$ simulations that are summarised in
Table~\ref{tab:sims}.  In  the  last  column we  show  the  $\sigma_8$
parameter for  the different  cosmological models. In  particular, for
the $f(R)$ models we have computed $\sigma_8$ from linear theory using
MG-CAMB \citep{zhao09a,hojjati11},  while for  the combined  models of
massive neutrinos  plus $f(R)$ we  joined together the  predictions of
CAMB for massive neutrinos with those of MG-CAMB.  For all simulations
we stored  34 full  snapshots for  a set of  redshifts that  allows to
construct without any  gap the lensing light-cones up  to $z_s=4$ that
are described below.
\begin{table*}
\begin{tabular}{lcccccccc}
Simulation Name & Gravity type  &  
$f_{R0} $ &
$m_{\nu }$ [eV] &
$\Omega _{\rm CDM}$ &
$\Omega _{\nu }$ &
$m^{p}_{\rm CDM}$ [M$_{\odot }/h$] &
$m^{p}_{\nu }$ [M$_{\odot }/h$] & $\sigma_8$ \\
\\
\hline
$\Lambda $CDM & GR & -- & 0 & 0.31345 & 0 & $8.1\times 10^{10}$  & 0 & $0.847$ \\
$fR4$ & $f(R)$  & $-1\times 10^{-4}$ & 0 & 0.31345 & 0 & $8.1\times 10^{10}$  & 0 & $0.967$ \\
$fR5$ & $f(R)$  & $-1\times 10^{-5}$ & 0 & 0.31345 &0  & $8.1\times 10^{10}$  & 0 & $0.903$ \\
$fR6$ & $f(R)$  & $-1\times 10^{-6}$ & 0 & 0.31345 & 0 & $8.1\times 10^{10}$  & 0 & $0.861$ \\
$fR4\_0.3eV$ & $f(R)$  & $-1\times 10^{-4}$ & 0.3 & 0.30630 & 0.00715 & $7.92\times 10^{10}$ & $1.85\times 10^{9}$ & $0.893$ \\
$fR5\_0.15eV$ & $f(R)$  & $-1\times 10^{-5}$ & 0.15 & 0.30987 & 0.00358 & $8.01\times 10^{10}$ & $9.25\times 10^{8}$ & $0.864$ \\
$fR5\_0.1eV$ & $f(R)$  & $-1\times 10^{-5}$ & 0.1 & 0.31107 & 0.00238 & $8.04\times 10^{10}$ & $6.16\times 10^{8}$ & $0.878$ \\
$fR6\_0.1eV$ & $f(R)$  & $-1\times 10^{-6}$ & 0.1 & 0.31107 & 0.00238 & $8.04\times 10^{10}$ & $6.16\times 10^{8}$ & $0.836$ \\
$fR6\_0.06eV$ & $f(R)$  & $-1\times 10^{-6}$ & 0.06 & 0.31202 & 0.00143 & $8.07\times 10^{10}$ & $3.7\times 10^{8}$  & $0.847$ \\
\hline
\end{tabular}
\caption{The subset of the \dustp  simulations considered in this work
  with  their   specific  parameters,   $f_{R0}$  represents   the  MG
  parameter,   $m_\nu$   and   $m^p_{\nu}$  the   neutrino   mass   in
  electron-volt and in $m_{\odot}/h$ as implemented in the simulation,
  $m^p_{\rm  CDM}$ Cold  Dark Matter  particle mass,  and $\Omega_{\rm
    CDM}$ and  $\Omega_{\nu}$ the $\mathrm{CDM}$ and  neutrino density
  parameters, respectively. In the last column of the table we display
  the  $\sigma_8$ parameter  at $z=0$  for the  different cosmological
  models computed from linear theory.} \label{tab:sims}
\end{table*}

\begin{figure*}
  \includegraphics[width=\hsize]{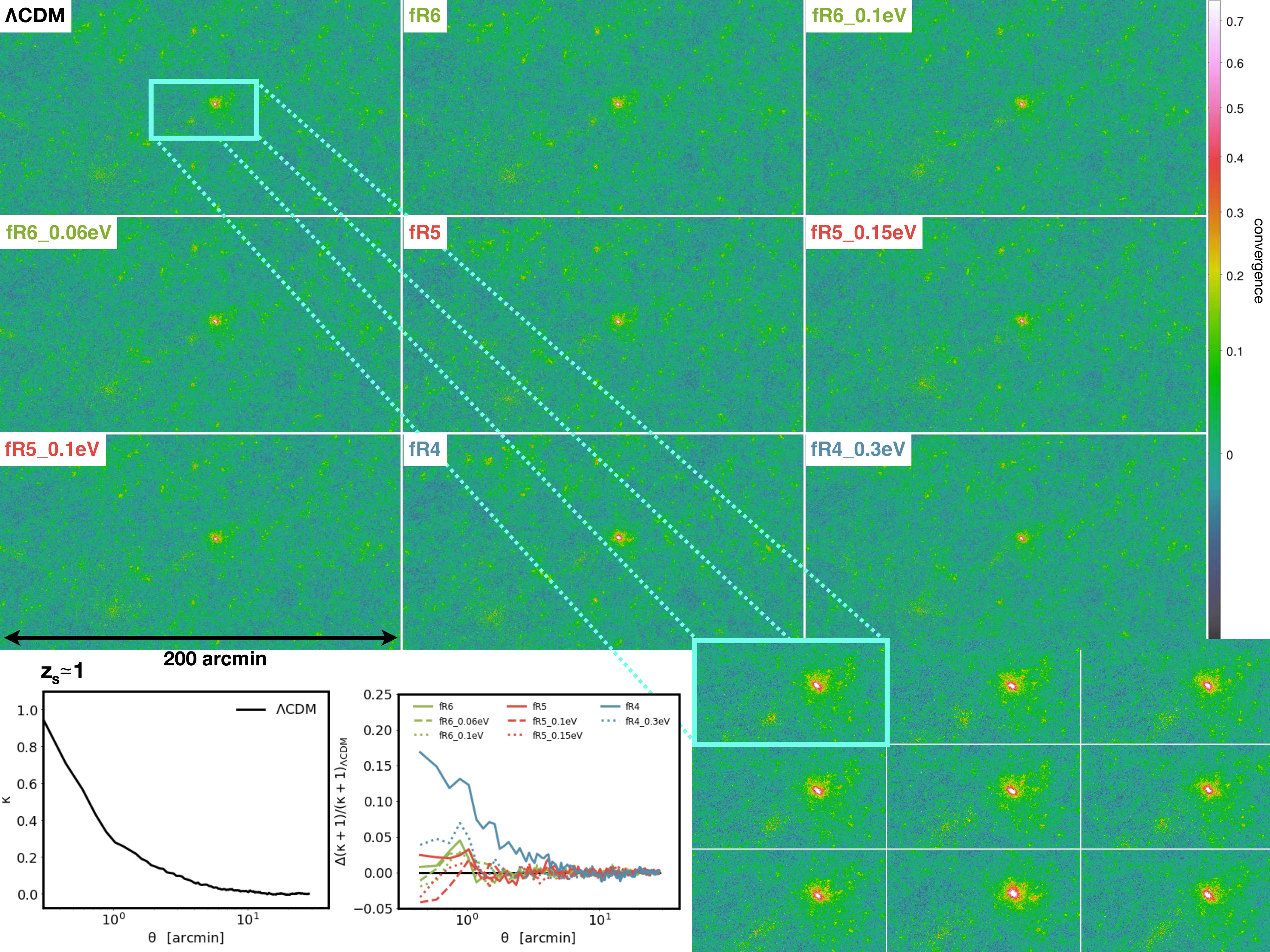}
  \caption{Convergence maps of a light-cone realisation for sources at
    redshift $z_s=1$. The various panels show the convergence maps for
    the  different   cosmological  simulations,  as  labeled   in  the
    panels. In the sub-panel we display  a zoom toward an area where a
    large  cluster  is located:  here  small  differences between  the
    different models  are more  visible. The two  plots in  the bottom
    left  part of  the figure  show:  the convergence  profile of  the
    cluster  in  the  \lcdm  model  and  the  relative  difference  of
    $(\kappa+1)$ between  the cluster  in the  various models  and the
    standard reference one. The color scale refers to the value of the
    convergence corresponding to each pixel of the maps.
    \label{FigConv}}
\end{figure*}

\subsection{The Halo Catalogues}
\label{halocatalogues}

For  all  simulations  we  have  identified  collapsed  $\mathrm{CDM}$
structures in each comoving snapshot  by means of a Friends-of-Friends
algorithm      \citep[FoF,][]{Davis_etal_1985}     run      on     the
$\mathrm{CDM}$\footnote{For  the simulations  containing also  massive
  neutrinos we decided  to follow the approach of not  linking them to
  the  collapsed haloes.}   particles with  linking length  $\lambda =
0.16  \times d$,  where  $d$ is  the  mean inter-particle  separation,
retaining  only   structures  with   more  than   $32$  $\mathrm{CDM}$
particles.  On  top of  such FoF  catalogues we  have run  the {\small
  SUBFIND}   algorithm   \citep[][]{Springel_etal_2001}  to   identify
gravitationally bound structures and  to associate standard quantities
such as the  virial mass $M_{200}$ and the virial  radius $R_{200}$ to
the main  substructure of each  FoF group.  The latter  quantities are
computed in the usual way by  growing spheres of radius $R$ around the
most-bound particle of  each main substructure enclosing  a total mass
$M$ until the relation
\begin{equation}
\frac{4}{3}\pi R_{200}^{3}\times 200 \times \rho _{crit} = M_{200}
\end{equation}
is   fulfilled   for   $R=R_{200}$  and   $M=M_{200}$,   where   $\rho
_{crit}\equiv 3H^{2}/8\pi G$ is the critical density of the universe.

\subsection{Past-Light-Cones}
\label{plc}

Several possible approaches  can be followed to  extract lensing light
cones from  large cosmological N-body simulations.   Recent efforts in
the  context  of  Modified  Gravity  simulations  have  employed  both
post-processing  reconstructions based  on  the slicing  of  a set  of
comoving particle snapshots \citep[as e.g.  in][]{shirasaki17} as well
as more  efficient on-the-fly algorithms  capable of storing  only the
projected matter density on a given field of view without resorting on
the            flat-sky            approximation            \citep[see
  e.g.][]{Barreira_etal_2016,Arnold_etal_2018}.     While   for    the
full-scale {\small DUSTGRAIN} simulations  the latter approach will be
also adopted  -- following  in particular  the same  implementation of
\citep{Arnold_etal_2018}  --  for the  present  analysis  of the  {\em
  pathfinder}  runs we  stick to  the former  procedure, and  for each
N-body  simulation we  build  the past-light-cones  using the  \mapsim
routine \citep{giocoli14}.   The particles from various  snapshots are
distributed  onto different  lens planes  according to  their comoving
distances with respect to the observer  and to whether they lie within
a defined aperture of the  field-of-view.  We use the particles stored
in $21$  different snapshots to construct  continuous past-light-cones
from  $z=0$  to $z=4$,  with  a  square sky  coverage  of  $5$ deg  by
side. Considering  the good time  resolution with which  the snapshots
have  been stored,  we are  able  to build  $27$ lens  planes for  the
projected  matter density  distribution.  In  \mapsim the  observer is
placed at the vertex of a pyramid whose square base is at the comoving
distance  corresponding  to $z=4$.   For  each  cosmological model  we
construct $256$  different light-cone realisations by  randomising the
various  comoving  cosmological  boxes  through  combinations  of  the
following  procedures:  {\em  i)}   changing  sign  of  the  cartesian
coordinates; {\em  ii)} redefining  the position  of the  observer and
{\em iii)} modifying  the order of the axes in  the coordinate system.
By construction,  these variations preserve the  clustering properties
of  the   particle  distribution   at  a  given   simulation  snapshot
\citep{roncarelli07}.     The   recent    improvements   of    \mapsim
\citep{giocoli17,castro17} give  us also the possibility  to store the
corresponding halo  and subhalo catalogues associated  with a specific
particle randomisation of the past light-cone.

The  \mapsim pipeline  allows  us the  construct  lensing planes  from
different simulation  snapshots, saving  for each  plane $l$,  on each
pixel,  with coordinate  indices  $(i,j)$, the  particle surface  mass
density $\Sigma$:
\begin{equation}
\Sigma_l(i,j) = \dfrac{\sum_k m_k}{A_{l}}\,,
\end{equation}
where $A_{l}$ represents the comoving pixel area of the $l$-lens plane
and $\sum_k  m_k$ the sum on  all particle masses associated  with the
pixel. Being  gravitational lensing sensitive to  the projected matter
density distribution along the line-of-sight,  onto each lens plane we
project all particles between two  defined comoving distances from the
observer; in  the simulations  with massive neutrinos  we consistently
account     also    for     this     component.      As    done     by
\citet{petri16,petri17,giocoli17,giocoli18a,  castro17}  we  construct
the convergence  map weighting the  lens planes by the  lensing kernel
and  assuming the  Born  approximation  \citep{bartelmann01}. In  this
respect, as  discussed in \citet{giocoli16a} and  \citet{castro17} the
Born approximation  is a very  good estimate of the  convergence power
spectrum and of the  Probability Distribution Function (hereafter PDF)
down to  scales of  a few  arcseconds. In  addition \citet{schaefer12}
have demonstrated,  by computing  an analytic  perturbative expansion,
that the Born approximation is an excellent estimation for weak cosmic
lensing down to very small scales ($l \geq 10^4$).

From $\Sigma_l$  we can write down  the convergence map $\kappa$  at a
given source redshift $z_s$ as:
\begin{equation}
\kappa = \sum_l \dfrac{\Sigma_l}{\Sigma_{\rm{crit},l,s}}\,,
\end{equation}
where $l$ varies over the different lens planes with the lens redshift
$z_l$ smaller  than $z_s$ and $\Sigma_{\rm{crit},l,s}$  represents the
critical  surface density  at  the  lens plane  $z_l$  for sources  at
redshift $z_s$ that can be read as:
\begin{equation}
\Sigma_{\rm{crit},l,s} \equiv \frac{c^2}{4 \pi G} \frac{D_l}{D_s
  D_{ls}}\,
\end{equation}
where $c$ indicates the speed of  light, $G$ the Newton's constant and
$D_l$, $D_s$ and  $D_{ls}$ are the angular  diameter distances between
observer-lens, observer-source and source-lens, respectively.

In Fig.~\ref{FigSk} we display a schematic representation of haloes in
a past-light-cone  from $z=0$ up to  $z=4$.  The left panel  shows all
haloes more  massive than $6  \times 10^{13} M_{\odot}/h$  in comoving
coordinates, colour  coded depending on  their mass.  The  right panel
displays the convergence map for $z_s=4$ with super-imposed the haloes
lying within the field-of-view.   Each constructed convergence map has
a square aperture of $5$ deg  by side and is resolved with $2048\times
2048$ pixels, which gives a  pixel angular resolution of approximately
$9$ arcsec.

The convergence maps for $z_s=1$  of the same light-cone randomisation
for  the   various  cosmological   models  are  displayed   in  Figure
\ref{FigConv}.  The  top left  panel refers  to the  \lcdm simulation.
Within  each map  we tag  the corresponding  cosmological model.   The
bottom right  sub-panel exhibits a zoom  toward the centre of  the map
where we  noticed the presence of  a massive galaxy cluster  with mass
$M_{200} = 1.14\times 10^{15}\;M_{\odot}/h$ at redshift $z=0.29$. This
value refers to  the mass enclosing 200 times the  critical density of
the  universe at  that  redshift  in the  \lcdm  model;  in the  other
cosmological models  the cluster mass  has a slightly  different value
well within $5\%$ except for $fR5\_0.15$eV in which it is about $10\%$
smaller and  for $fR4$  in which  it is more  than $40\%$  larger. The
different value of  $M_{200}$ in the various models  indicates that we
are probably  observing the  same structure in  different evolutionary
phases:  in particular  in $fR5\_0.15$eV  ($fR4$) the  system is  less
(more) evolved,  which will result in  different structural properties
like  e.g.    the  concentration  parameter   \citep{giocoli07}.   The
sub-panels  on  the  bottom  left  part  of  the  figure  display  the
convergence properties  of the  cluster present at  the centre  of the
maps. The left one shows the convergence $\kappa$ profile in the \lcdm
simulation while the one on the right displays the relative difference
between $(\kappa+1)$ profile on the  same system in the various models
with respect to  the \lcdm measurements. From this figure  it is quite
evident  that  the cluster  in  the  $fR4$  model  has a  very  picked
convergence  profile  that rises  $15\%$  more  than in  the  standard
cosmology,  resulting   in  a  much  higher   concentration  parameter
\citep{barreira17}.

\begin{figure}
  \includegraphics[width=\hsize]{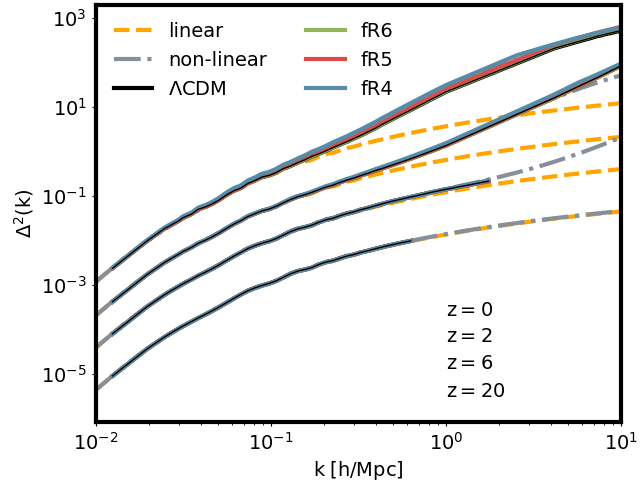}
  \caption{The dimensionless  matter power  spectra at  four different
    redshifts, $z=0$,  $2$, $6$,  $20$ from  top to  bottom. Different
    colors  display  the  results  for the  various  Modified  Gravity
    models: $fR6$ (green), $fR5$ (red)  and $fR4$ (blue). Black curves
    display the prediction  for the \lcdm simulation  while the orange
    dashed  and   dark-gray  dot-dashed  ones  show   the  linear  and
    non-linear matter  power spectrum  at the  corresponding redshifts
    from CAMB  \citep{camb}. For the non-linear  matter power spectrum
    we      have      considered       the      implementation      by
    \citet{takahashi12}.\label{FigPk3D}}
\end{figure}

\subsection{The Matter Power Spectra}
\label{matterpk}

The total  matter power  spectrum represents a  challenging statistics
for future  wide field  surveys to  discriminate between  standard and
non-standard models.   The large amount  of data expected  from future
wide field surveys, and the great number of available sources for weak
lensing   measurements    expected   for   the    ESA-Mission   Euclid
\citep{euclidredbook}, will  offer the  possibility to  constrain with
unprecedented accuracy  the Dark Energy  equation of state,  the total
neutrino  mass, and  to detect  possible  deviations from  GR by  {\em
  tomographically}  measuring  the  growth  rate of  structures  as  a
function of the weak lensing source redshift.

\begin{figure*}
  \includegraphics[width=0.495\hsize]{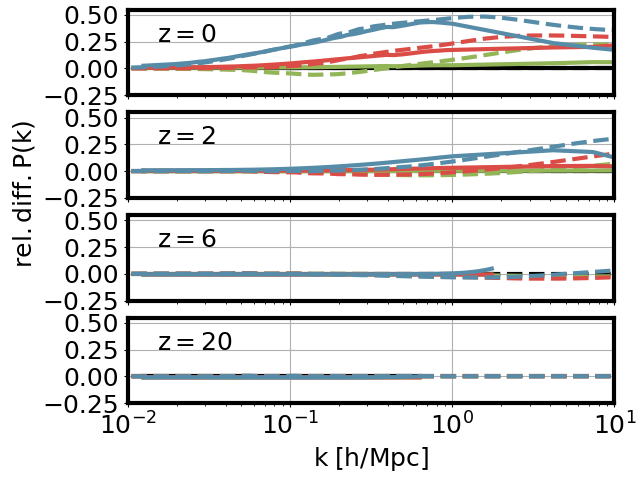}
  \includegraphics[width=0.495\hsize]{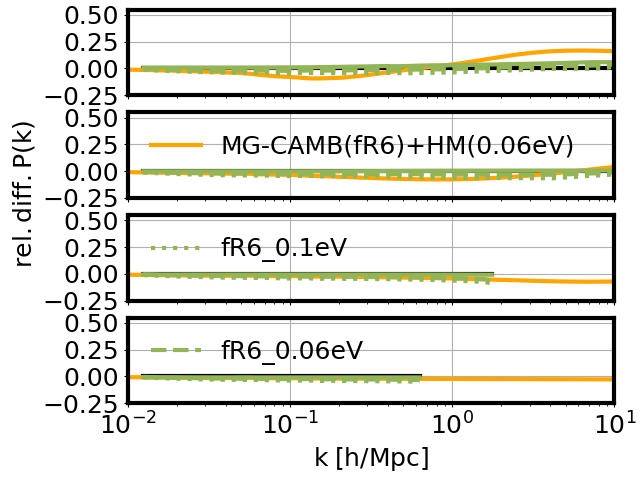}
  \includegraphics[width=0.495\hsize]{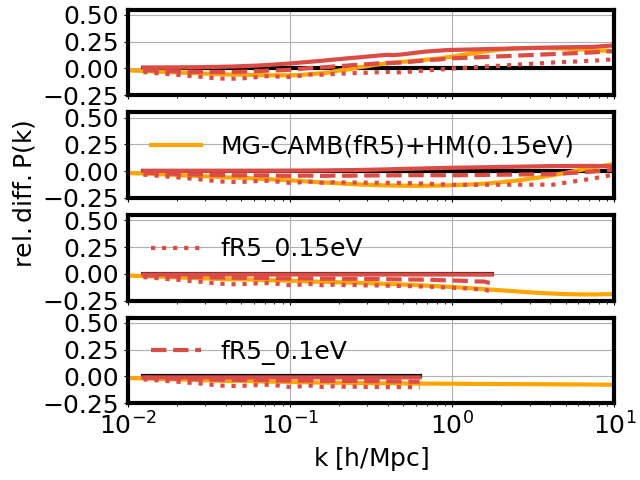}
  \includegraphics[width=0.495\hsize]{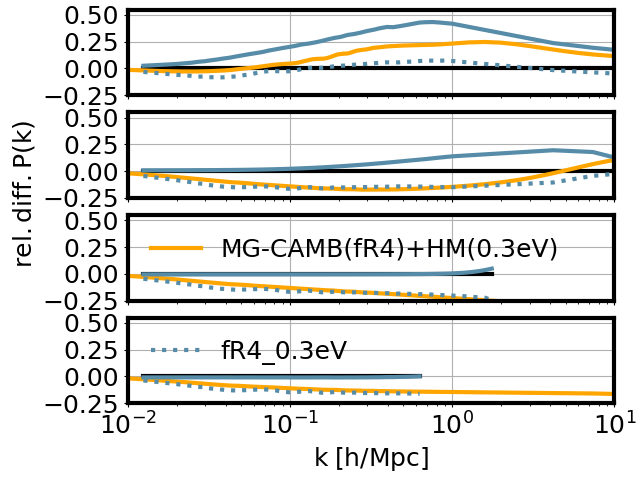}        
  \caption{Relative differences between the  matter power spectra of a
    given cosmological model with respect to the \lcdm one. The panels
    from top  to bottom display  the results  for $z=0$, $2$,  $6$ and
    $20$,  respectively.  The  top  left panels  display the  relative
    differences of the pure Modify Gravity models, while the others --
    for each MG case, green, red  and blue for $fR6$, $fR5$ and $fR4$,
    respectively -- show the differences when also a massive neutrinos
    component is included.\label{FigRel3D}}
\end{figure*}

For each of our cosmological simulations and stored snapshots, we have
computed the  total matter power  spectrum by determining  the density
field  on a  cubic Cartesian  grid with  twice the  resolution of  the
Particle  Mesh (hereafter  PM) grid  used for  the N-body  integration
(i.e.    $768^3$   grid   nodes)    through   a   Cloud-in-Cell   mass
assignment. This procedure provides  a determination of the non-linear
matter power  spectrum up  to the  Nyquist frequency  of the  PM grid,
corresponding to $k_{Ny}  = \pi N/L \approx  3.2\;h/$Mpc. The obtained
power spectrum is then truncated at  the $k$-mode where the shot noise
reaches $20\%$ of the measured power.

Figure \ref{FigPk3D}  shows the dimensionless matter  power spectra at
four  different redshifts  for  the  pure MG  and  the standard  \lcdm
models.  From  top to bottom we  display redshift $z=0$, $2$,  $6$ and
$20$, respectively, which present a constant shift equal to the square
of the linear growth factor.  The black lines display the measurements
for  the \lcdm  simulation; green,  red and  blue lines  refer to  the
analyses  for the  Modified  Gravity models  $fR6$,  $fR5$ and  $fR4$,
respectively.  The  orange dashed and the  dark-gray dot-dashed curves
show  the  predictions  from  linear  and  non-linear  theory  at  the
corresponding redshifts computed using \textsc{camb} \citep{camb}, for
the same cosmological  parameters of the \lcdm  run. Specifically, for
the   non-linear   matter  power   spectrum   we   have  adopted   the
parametrisation by \citet{takahashi12} which  is in agreement with the
\lcdm measurements within  few percents (apart for  the particle noise
contribution appearing above $k \gtrsim 5$ $h$/Mpc for $z=2$), as well
as  the halo  model predictions  by \citet{mead15b}  not displayed  to
avoid overcrowding the figure.

\begin{figure}
  \includegraphics[width=0.96\hsize]{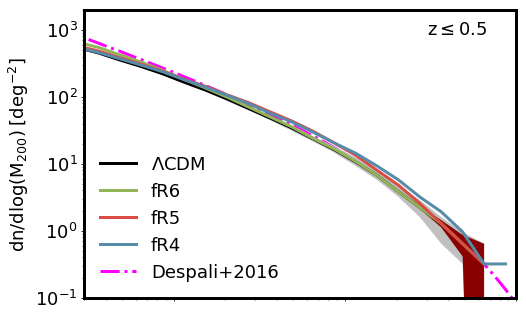}
  \includegraphics[width=0.96\hsize]{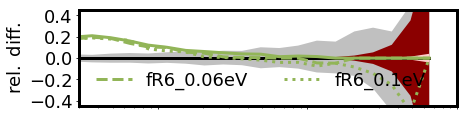}
  \includegraphics[width=0.96\hsize]{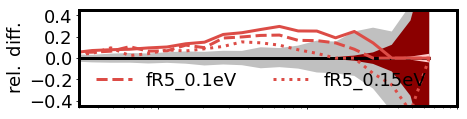}
  \includegraphics[width=\hsize]{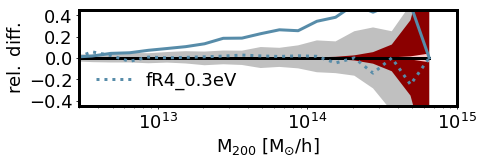}
  \caption{Halo  mass  function  per  unit square  degree  within  the
    past-light-cone up to $z=0.5$, in the various cosmological models.
    The  curves  display the  median  counts  in the  $256$  different
    light-cone  realisations, the  gray  shaded  area surrounding  the
    \lcdm measurements define the first and the third quartiles of the
    distribution.   The   red  and  pink  regions   mark  the  Poisson
    uncertainties of the halo counts within $25$ and $15,000$ (angular
    size  of  the  Euclid wide  survey  \citep{euclidredbook})  square
    degrees.  The  dot-dashed magenta  curve shows the  prediction for
    the  \lcdm   model  computed  using  the   \citet{despali16}  mass
    function. \label{figlowzmf}}
\end{figure}

\begin{figure}
  \includegraphics[width=0.96\hsize]{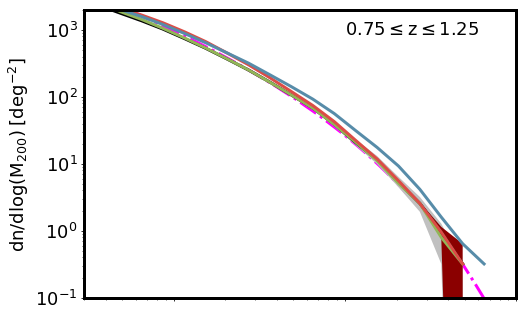}
  \includegraphics[width=0.96\hsize]{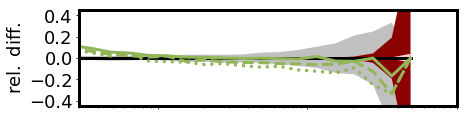}
  \includegraphics[width=0.96\hsize]{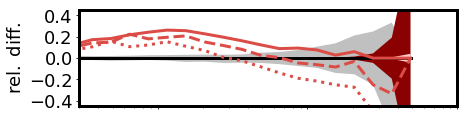}
  \includegraphics[width=\hsize]{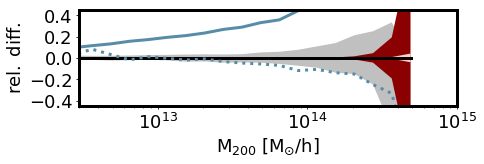}
  \caption{As Fig.~\ref{figlowzmf} but for haloes between $z=0.75$ and
    $z=1.25$.\label{highzmf}}
\end{figure}

The relative differences between the  various models and the \lcdm one
are displayed in  Fig.~\ref{FigRel3D}.  In the top left  panel we show
the differences for  the pure MG models, together  with the non-linear
predictions from MG-CAMB displayed by the corresponding dashed curves.
In these panels  we notice that MG-CAMB predicts quite  well the large
scale power spectra behaviour up to  $k\sim 1$ Mpc/$h$ while the small
scale  trends  stay above  the  simulation  measurements.  This  is  a
well-known  feature  related  to  the onset  of  the  {\em  Chameleon}
screening mechanism at non-linear scales, that was already noticed and
highlighted in the very first simulations of $f(R)$ gravity \citep[see
  e.g.  Fig.~2 of ][]{Oyaizu_etal_2008}  and subsequently confirmed by
several other studies  \citep[see e.g.][]{Zhao_Li_Koyama_2011a}.  More
quantitatively,  from  the  figure  we notice  that  the  $fR4$  model
displays at $z=0$ the typical  enhancement of $\sim 50\%$ with respect
to     the      standard     model     at      $k\approx     6\;h/$Mpc
\citep[][]{Oyaizu_etal_2008}, at  $z=2$ the relative  difference moves
toward smaller scales,  $k\approx 60\;h/$Mpc, decreasing approximately
by  a factor  of two.   The $fR5$  model for  all redshifts  $z\neq 0$
remains close  to the  standard $\mathrm{\Lambda  CDM}$, while  at the
present time it shows an enhancement that monotonically increases from
$k=0.1  h/$Mpc reaching  $25\%$ at  $k=10\;h/$Mpc.  The  $fR6$ at  all
redshifts and scales remains very close to the standard model.

In the other three panels we include also the models which account for
the combined effects of MG and  massive neutrinos, as indicated in the
corresponding labels.  In those panels we also display in solid orange
some combined  predictions of  MG-CAMB and  non-linear CAMB  using the
halo  model   by  \citet{mead16}  for  the   massive  neutrinos  runs;
specifically, in those cases we are  assuming to be able to simply add
on  top of  each other  the relative  contribution of  MG-CAMB to  the
\citet{mead16}  predictions. At  high redshifts,  where the  effect of
$f(R)$ gravity is  still very weak even for the  largest values of the
scalar   amplitude  $f_{R0}$   under   consideration,  this   combined
prediction follows  quite well the  fully non-linear result  (see e.g.
the $fR4\_0.3$eV case down to  $z=2$).  However, at lower redshifts we
can observe  that the simple  sum of the two  non-standard predictions
fails in capturing the real evolution of the matter power spectrum, in
particular  at   low  redshifts,   mainly  toward  large   $k$.   This
demonstrates  the   need  to  employ  full   numerical  solutions,  or
alternatively some faster approximate  methods such as those developed
by \citet{Wright_Winther_Koyama_2017},  to investigate  and accurately
predict   non-linear  observables   in  these   combined  cosmological
scenarios.  The  main qualitative statement arising  from the analysis
of Fig.~\ref{FigRel3D} is  that in all simulations  where also massive
neutrinos  are  included the  relative  difference  of the  of  $f(R)$
gravity power spectrum with respect to  GR is suppressed.  At the same
time it  is worth to notice  that the relative contribution  of MG and
massive  neutrinos to  the  total matter  power  spectrum is  redshift
dependent.  In  particular, focusing  our attention on  the simulation
$fR4\_0.3eV$, we  see that at  high redshifts neutrinos tend  to lower
the power on  large scales by about $15-20\%$, while  at low redshifts
the MG contribution acts on small scales producing at the present time
a total matter  power spectrum very close (within few  percent) to the
\lcdm one.

To summarise,  in this  section we have  noticed that  $f(R)$ Modified
Gravity and  massive neutrinos  affect the matter  density fluctuation
field at similar scales but with  opposite effects and with a slightly
different redshift  evolution.  While  Modified Gravity results  in an
enhancement of  the matter power  spectrum at scales smaller  than the
associated Compton wavelength, massive  neutrinos tend to suppress the
matter  power spectrum  on a  similar range  of scales  so that  their
combined  effect  may  result  in a  significant  weakening  of  their
individual  characteristic footprints.   These results  are all  fully
consistent        with       previous        findings       \citep[see
  e.g.][]{Baldi_etal_2014,Wright_Winther_Koyama_2017}

Although  providing  useful  insights  on the  effects  of  these  two
phenomena, the  comoving three-dimensional matter power  spectrum that
we   have   discussed   so   far  is   not   a   directly   observable
quantity. Present and future wide-field cosmological surveys typically
measure projected  statistics of  the matter density  distribution and
clustering properties of three-dimensional  galaxy catalogues within a
past  light-cone.  Therefore,  in  order to  mimic real  observational
experiments, we  have constructed a  set of past-light-cones  for each
cosmological simulation, and in the next sections we will focus on the
statistical analysis  of weak  lensing and cluster  counts observables
within these  light-cones, thereby extending the  investigation of the
degeneracy from  the comoving matter distribution  studied in previous
works to the more realistic case of weak lensing statistics.

\section{Results}
\label{results}

\begin{figure}
  \hspace{0.03\hsize}\includegraphics[width=0.97\hsize]{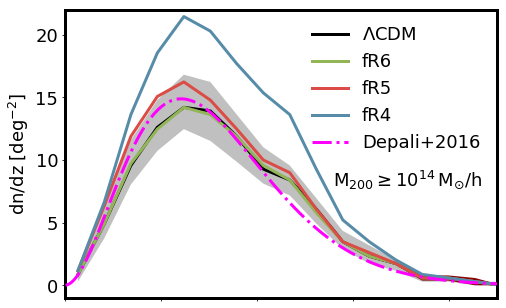}
  \includegraphics[width=\hsize]{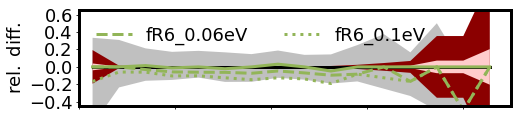}
  \includegraphics[width=\hsize]{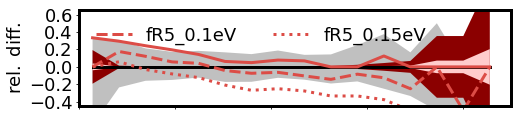}
  \includegraphics[width=\hsize]{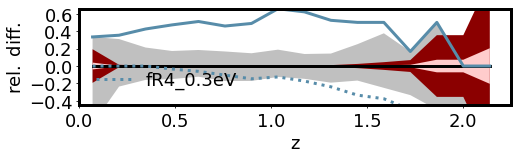}    
  \caption{Cluster -- haloes  more massive then $10^{14}\,M_{\odot}/h$
    -- redshift distribution for the various cosmological models.  The
    various  curves  display  the   median  counts  in  the  different
    light-cone realisations while the  shaded grey area bracketing the
    \lcdm measurements  defines the first  and the third  quartiles of
    the  distribution. The  red and  the  pink area  mark the  Poisson
    uncertainties  of  the  counts  within $25$  and  $15,000$  square
    degrees. The  three sub-panels display the  relative difference of
    the counts in the different MG models, with or without the massive
    neutrino components,  with respect  to the  $\mathrm{\Lambda CDM}$
    ones.\label{figdndz}}
\end{figure}

In this  section we present  the statistical properties of  haloes and
projected   particle    density   distribution    within   constructed
past-light-cones. These analyses will help us shed more light on 'what
and where  to look at' to  better disentangle the various  models with
respect to the standard \lcdm reference scenario.

\subsection{Halo Mass Functions and Redshift Distribution}
\label{hmf}

\begin{figure}
  \includegraphics[width=\hsize]{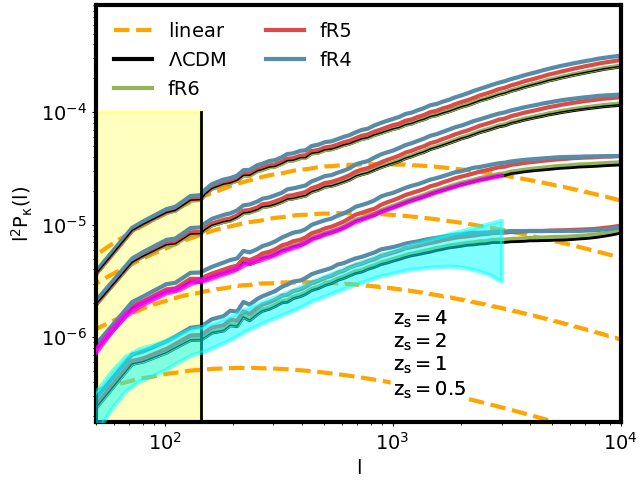}
  \caption{Convergence  power  spectra  at four  different  redshifts:
    $z_s=4$, $2$,  $1$ $0.5$  from top  to bottom,  respectively.  The
    black  curves   display  the   average  measurements   from  $256$
    light-cone random realisations for the \lcdm model. Green, red and
    blue curves show  the measurements for the $fR6$,  $fR5$ and $fR4$
    models, respectively; orange dashed curves refer to the prediction
    using linear matter  power spectrum for the  \lcdm cosmology using
    CAMB.\label{figpkl} The black vertical line marks the angular mode
    corresponding to half  field of view $l_{\rm  half}=144$. The pink
    and cyan shaded area illustrate the observational uncertainties --
    up to $l=3000$ -- associated to  the power spectra for a survey of
    $15,000$ and $154$ square degrees  considering a number density of
    galaxies of  $8$ and  $33$ per  square arcmin,  with corresponding
    average source redshift of $z_s=1$ and $z_s=0.5$, respectively.}
\end{figure}

The effect  of MG  on the  dynamical evolution  of the  matter density
field results in different halo formation epochs and number density of
collapsed systems.   In this respect  it is important to  mention that
cluster counts represent a promising statistics for future surveys, as
their  number  density  and  redshift distributions  are  expected  to
significantly  improve   the  constraints  on   standard  cosmological
parameters coming  from other  probes, as  e.g. galaxy  clustering and
weak   lensing   \citep[see  e.g.][for   the   case   of  the   Euclid
  survey]{sartoris16}.

In Figure \ref{figlowzmf} we show the median 2D halo mass function per
square  degree, over  $256$ realisations,  obtained by  populating the
past light-cones with our halo catalogues up to $z=0.5$. The top panel
shows the measurements  for the \lcdm cosmology and the  three pure MG
models,  in  the   three  bottom  sub-panels  we   show  the  relative
differences of  the corresponding MG  models with and  without massive
neutrinos  with  respect  to  the  \lcdm  case.   The  shaded  regions
bracketing the black  lines mark the first and the  third quartiles of
the measurements  in the  \lcdm simulation  computed in  the different
light-cone  randomizations,  the  red  and pink  regions  display  the
Poisson  uncertainties in  halo  counts in  $25$  and $15,000$  square
degrees,  respectively;   the  latter  represents  the   area  of  the
ESA-Euclid wide  survey \citep{euclidredbook}. The  dot-dashed magenta
curve shows  the theoretical  prediction by \citet{despali16}  for the
corresponding mass overdensity definition,  which describes quite well
the numerical simulation results for the corresponding \lcdm cosmology
down  to approximately  $10^{13}\;M_{\odot}/h$,  which corresponds  to
haloes resolved with  at least $125$ dark-matter  particles.  From the
first  sub-panel  we notice  that  for  masses above  $M_{200}>2\times
10^{13}M_{\odot}/h$ all $fR6$  models are quite close,  and within the
grey region, to  the \lcdm cosmology, sharing a very  similar value of
$\sigma_8$ within few percents; however at smaller masses they tend to
predict significantly more haloes, with an increase of about $20-30\%$
for $M_{200}<10^{13}M_{\odot}/h$.   Differently from this  first case,
the second  sub-panel, referring  to the  $fR5$ cosmologies,  does not
show a  monotonic trend: while for  small and large masses  the counts
are     close    to     those    in     the    \lcdm     model,    for
$10^{13}M_{\odot}/h<M_{200}<10^{14}M_{\odot}/h$ the  $fR5$ models show
$10-35\%$ more  haloes with a  difference decreasing as a  function of
the  total  neutrino mass.   In  the  last  sub-panel we  display  the
measurements  in the  $fR4$ models;  while the  pure $fR4$  simulation
monotonically shows more haloes than the \lcdm case, consistently with
the larger  value of  $\sigma_8$; the  $fR4\_0.3$eV case  remains very
close to  the standard cosmology,  as expected  due to the  well known
degeneracy  between  these  two  values of  $f_{R0}$  and  $m_{\nu  }$
\citep[][]{Baldi_etal_2014,hagstotz18}.   In  general we  notice  that
while MG can  have qualitatively different effects  for different halo
masses depending on the value of the $f_{R0}$ parameter, the effect of
a massive  neutrino component is  always an increasing  suppression of
the halo counts  for increasing mass, with magnitude  that is stronger
for larger total neutrino masses.

As  discussed in  \citet{sartoris16},  future wide  field surveys  are
expected to provide important information also on the cluster and halo
counts at  higher redshifts. For  the Euclid  wide survey most  of the
clusters are expected to have a redshift between $0.6<z<1.2$, and even
larger.

In  order to  understand the  effect of  MG with  and without  massive
neutrinos at these higher redshifts,  in Fig. \ref{highzmf} we display
the halo mass function per  unit square degree for $0.75<z<1.25$.  The
relative trends  with respect  to \lcdm and  to lower  redshift counts
allows us  to better  trace the growth  of structures  in non-standard
models.  The data in the panels are analogous to Fig.~\ref{figlowzmf}.
From Fig.~\ref{highzmf} we notice that  in particular the $fR6$ models
show a  lower excess of low  mass haloes with respect  to the standard
cosmology  as compared  to  the lower  redshift  observations of  Fig.
\ref{figlowzmf}.  In  this redshift  bin the  low-mass systems  of the
$fR5$ models are more numerous by  about $20\%$ while the abundance of
more massive cluster-sized objects is  again consistent with the \lcdm
expectation.  In  the last  sub-panel we  show the  case of  the $fR4$
cosmologies: the model without massive  neutrinos has more haloes than
\lcdm, reaching a difference of  about $80\%$ for cluster-size haloes,
while the model featuring a $0.3$ eV neutrino mass is again very close
to  the standard  cosmology also  at these  redshifts.  These  results
clearly confirm the strong  degeneracy between $f(R)$ modified gravity
and massive neutrinos in the  non-linear regime of structure formation
that  has  been  first  pointed  out  by  \citet{Baldi_etal_2014}  and
subsequently     confirmed     by     other     studies     \citep[see
  e.g.][]{mead16,Bellomo_etal_2017,Wright_Winther_Koyama_2017,peel18}.

Next generation  of space missions,  like the Euclid ESA  mission, are
expected to use  cluster counts as a  complementary cosmological probe
to   weak   lensing   and   galaxy  clustering.    As   discussed   in
\citet{sartoris16}, photometric  identification of galaxy  clusters is
expected to deliver a catalogue of systems with $S/N>3$ with a minimum
mass of approximately  $M_{\rm min}\approx 10^{14}M_{\odot}/h$, almost
independently  of  redshift up  to  $z=2$.   In Fig.~\ref{figdndz}  we
display  the median  cluster  redshift distribution  in the  different
light-cone realisations.  The  black curve shows the  median counts in
the \lcdm past-light-cones and the shaded gray area brackets the first
and the third quartiles of the  distribution. The red and pink regions
display the Poisson  uncertainties of the cluster counts  for a survey
of $25$ and  $15,000$ square degrees.  Colours and line  styles are as
in the  previous figures, and  as before the dot-dashed  magenta curve
corresponds  to  the  prediction   by  \citet{despali16}.   The  three
sub-panels show  the relative  difference of the  various cosmological
models with respect to the standard one.  We may notice that all $fR6$
and $fR5$ models show a  cluster redshift distribution consistent with
\lcdm  well  within  the  sample  variance  except  for  the  case  of
$fR5\_0.15$eV that  presents about  $20\%$ fewer clusters  toward high
redshifts.   The  $fR4$  cosmology without  massive  neutrinos,  shows
approximately $40\%$  more cluster  than \lcdm,  while for  a neutrino
mass of $0.3$ eV the  cluster redshift distribution appears similar to
that of the  $fR5\_0.15$eV cosmology.  This is another  example of the
observational  degeneracy  we  aim  to investigate  with  the  {\small
  DUSTGRAIN} simulations: different values  of the total neutrino mass
may be inferred from the same  data by changing the assumptions on (or
even  just by  adjusting he  parameters of)  the underlying  theory of
gravity.  A more  detailed characterisation of the  halo mass function
in combined $f(R)$ and massive  neutrinos cosmologies -- calibrated on
the  same  \dustp simulations  described  in  this  work --  has  been
recently presented in a companion paper by \citet{hagstotz18}.

\begin{figure*}
  \includegraphics[width=0.49\hsize]{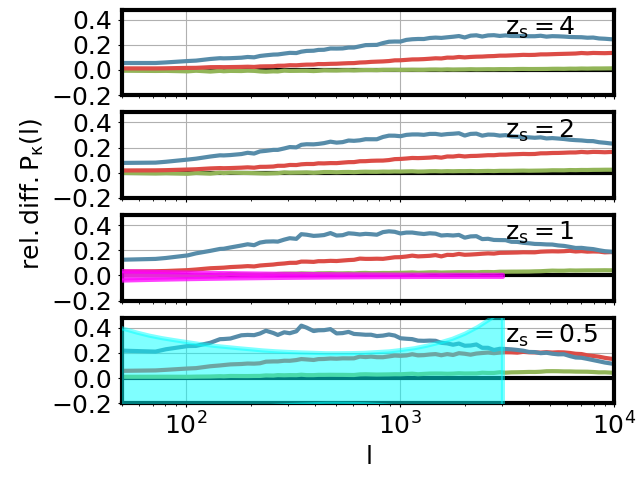}
  \includegraphics[width=0.49\hsize]{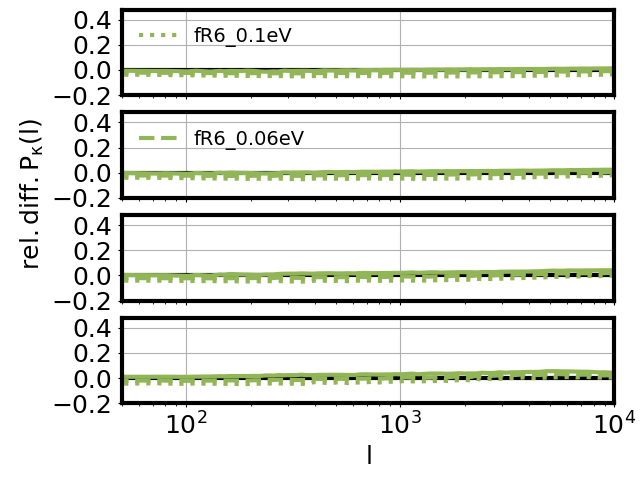}
  \includegraphics[width=0.49\hsize]{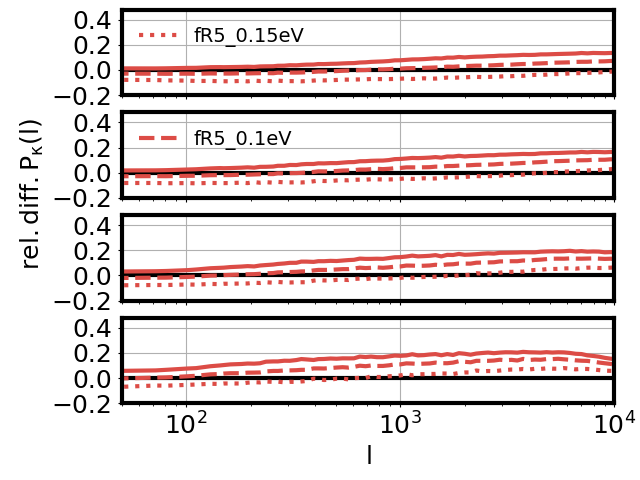}
  \includegraphics[width=0.49\hsize]{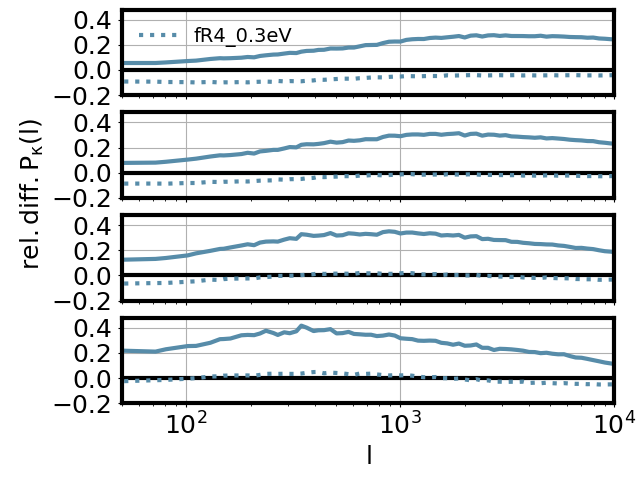}
  \caption{Relative difference  between the convergence  power spectra
    at four  considered source redshifts for  the various cosmological
    models with respect  to the \lcdm one.  In each  panel from top to
    bottom  we display  the  measurements for  $z_s=4$,  $2$, $1$  and
    $0.5$, respectively.   As in Fig.~\ref{FigRel3D}, in  the top left
    panel we display the relative  differences for the pure MG models,
    while in  the others we show  the MG with the  corresponding total
    massive neutrino components.\label{Dfigpkl}}
\end{figure*}

\subsection{Properties of the Convergence Maps}
\label{convergence}

The shape measurement of a large sample of background sources is known
to provide  an almost unbiased estimate  of the shear field  caused by
the interposed matter density distribution along the line-of-sight. In
the weak lensing regime the ellipticity  of a galaxy $\epsilon$ can be
written as \citep{seitz97}:
\begin{equation}
\epsilon \approx \epsilon^s + \frac{\gamma}{1-\kappa}\,,
\label{eqshape1}
\end{equation}
where $\epsilon^s$ represents the  intrinsic ellipticity in the source
plane and $\gamma=(\gamma_1,\gamma_2)$  is the value of  the shear; by
defining the  lensing potential $\psi$  as the projected  potential of
the three-dimensional matter density distribution \citep{kilbinger14},
we can write the components of $\gamma$ as
\begin{equation}
\gamma_1 = \frac{1}{2}(\partial_1 \partial_1 - \partial_2 \partial_2) \psi\,,\;\;
\gamma_1 = \partial_1 \partial_2 \psi\,,
\end{equation}
as well as for the convergence
\begin{equation}
\kappa = \frac{1}{2}(\partial_1 \partial_1 + \partial_2 \partial_2) \psi \,,
\end{equation}
where $\partial_i$  represents the partial derivative  with respect to
the $i$th-component  in the plane  of the  sky.  Averaging on  a large
sample of galaxies, and considering  that the intrinsic ellipticity of
the  source galaxies  has  no preferred  orientation,  we can  rewrite
eq.~(\ref{eqshape1}) as:
\begin{equation}
\langle \epsilon \rangle \approx g\,,
\end{equation}
where $g\equiv\gamma/(1-\kappa)$  represents the reduced  shear.  From
the  measured  reduced  shear   field  $g$  then  suitable  algorithms
\citep[like  e.g.   the  KSB:][]{kaiser95}  are used  to  recover  the
convergence map.  The weak lensing  convergence field is the result of
inhomogeneities in the large scale  matter distribution along the line
of sight to distant sources.

The convergence power spectrum, to first order, can be expressed as an
integral of the three-dimensional  matter power spectrum computed from
the observer looking at the  past-light-cone from the present epoch up
to a given source redshift \citep{bartelmann01}. In this approximation
it is assumed  that the light rays travel along  unperturbed paths and
all terms  higher than  first order  in convergence  and shear  can be
ignored.  Defining $f(w)$ as the angular radial function, that depends
on  the comoving  radial coordinate  $w$  given the  curvature of  the
universe,  we can  write the  convergence power  spectrum for  a given
source redshift $z_s$ -- with  a corresponding radial coordinate $w_s$
-- as:
\begin{equation}
P_{\kappa}(l) = \dfrac{9 H_0^4 \Omega_m^2}{4 c^4} \int_0^{w_s(z_s)}
\dfrac{f^2(w_s-w)}{f^2(w_s)a^2(w)} P_{\rm \delta}\left(\dfrac{l}{f(w)},w\right) \, \mathrm{d}w.
\label{eqborn}
\end{equation}
Analogously, from  the constructed  effective convergence maps  we can
compute the corresponding power spectrum as:
\begin{equation}
 \langle \hat{\kappa}(\mathbf{l}) \hat{\kappa^*}(\mathbf{l}') \rangle = {4 \pi^2}
 \delta_D(\mathbf{l}-\mathbf{l}') P_{\kappa}(l),\label{pkmapeq}
\end{equation}
where  $\delta_D^{(2)}$  represents  a  Dirac delta  function  in  two
dimensions.

\begin{figure*}
  \includegraphics[width=0.49\hsize]{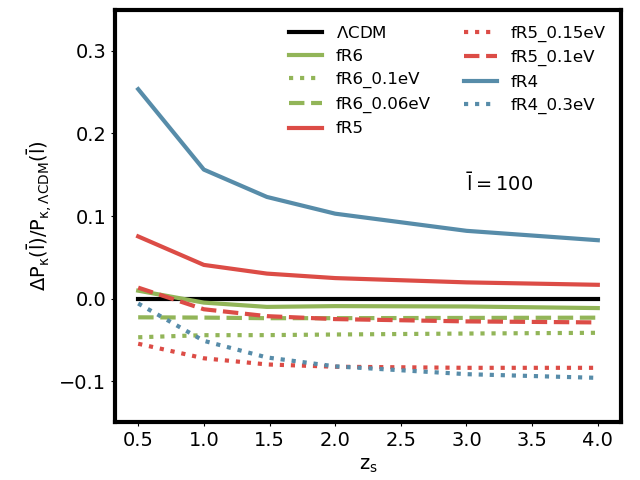}
  \includegraphics[width=0.49\hsize]{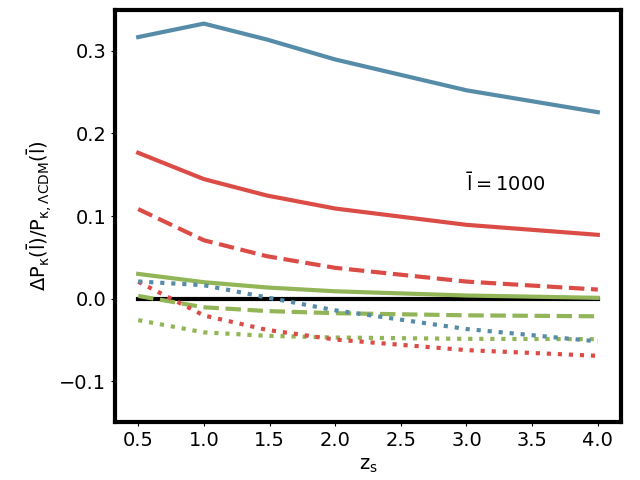}
  \caption{Ratio  between the  convergence power  spectra measured  at
    $l=100$  (left  panel)  and  $l=1000$ (right  panel)  between  the
    different non-standard models  and the \lcdm one as  a function of
    the source redshift $z_s$.\label{figlfixed}}
\end{figure*}

In Fig.~\ref{figpkl} we display the  convergence power spectra at four
different source redshifts:  $z_s=4$, $2$, $1$ and $0.5$,  from top to
bottom  respectively.    The  black   curve  represents   the  average
measurements over  the $256$ different light-cone  realisations in the
\lcdm  model.   The  black  vertical   line  marks  the  angular  mode
corresponding to half field of view $l_{\rm half}=144$. Our light-cone
simulations  mainly model  non-linear scales  of the  projected matter
density field,  predictions on larger  angular modes can  be addressed
using linear theory parametrisation \citep{schmidt08}.  The green, red
and  blue  curves  display  the  average  measurements,  at  the  same
corresponding redshifts, in the  three modified gravity models: $fR6$,
$fR5$  and $fR4$  respectively.   The orange  dashed  curves show  the
prediction from eq.~(\ref{eqborn}) using  the theoretical linear power
spectrum  for  the  \lcdm  cosmology.   The  pink  and  cyan  regions,
bracketing the  \lcdm predictions at  $z_s=0.5$ and $z_s=1$,  show the
observational  uncertainties  associated  to  the  shape  measurements
\citep{refregier04,lin15a,matilla16,shan17}, for  a number  density of
$8$ and $33$ galaxies per square arcminutes, mimicking -- respectively
-- a   ground-based   experiment   like  CFHTLens   \citep[$154$   sq.
  deg.,][]{kilbinger13}  and   a  future  space  survey   like  Euclid
\citep[$15,000$  sq.   deg.,][]{euclidredbook}.   The two  wide  field
experiments have  an average  source redshift  distribution consistent
with the numbers  considered above and typically  peak around redshift
$z=0.5$  and  $z=1$,  for  the ground-  and  space-based  experiments,
respectively.  Neglecting non-Gaussian corrections we have defined the
uncertainties   associated  with   the   convergence  power   spectrum
$P_{\kappa}(l)$ \citep{refregier04,huterer02} as
\begin{equation}
  \Delta P_{\kappa}(l) = \sqrt{\dfrac{2}{2(l+1) f_{\rm sky}}} \left(
  P_{\kappa}(l) + \dfrac{\sigma_{\epsilon}^2}{2 n_g}\right)\,,
\end{equation}
where $f_{\rm  sky}$ represents the fraction  of the sky covered  by a
given   survey,   $n_g$   the   number   density   of   galaxies   and
$\sigma_{\epsilon}=0.25$ the  rms of the intrinsic  ellipticity of the
sources.

\begin{figure*}
  \includegraphics[width=0.49\hsize]{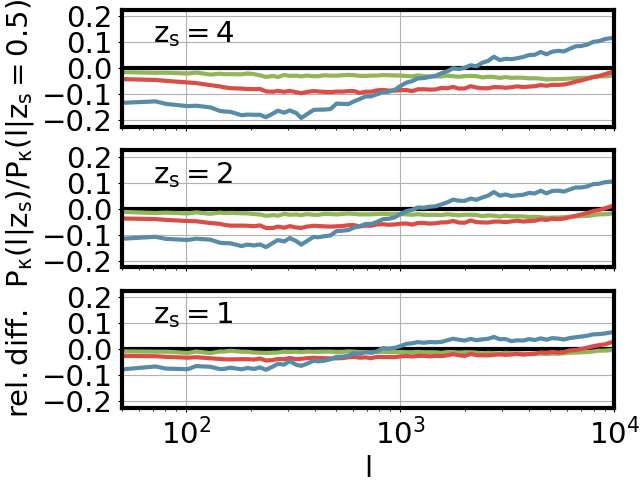}
  \includegraphics[width=0.49\hsize]{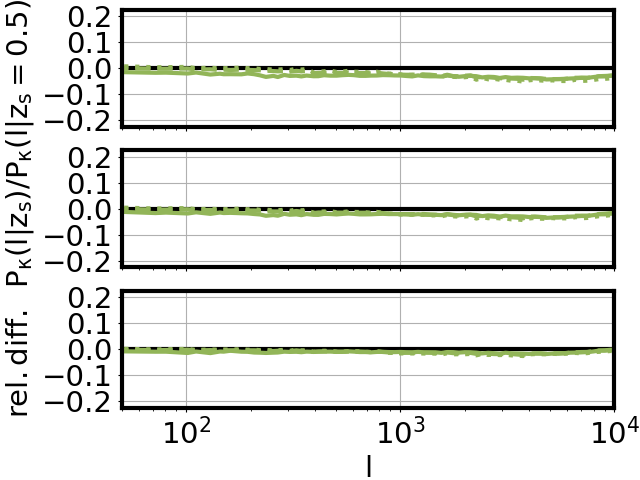}
  \includegraphics[width=0.49\hsize]{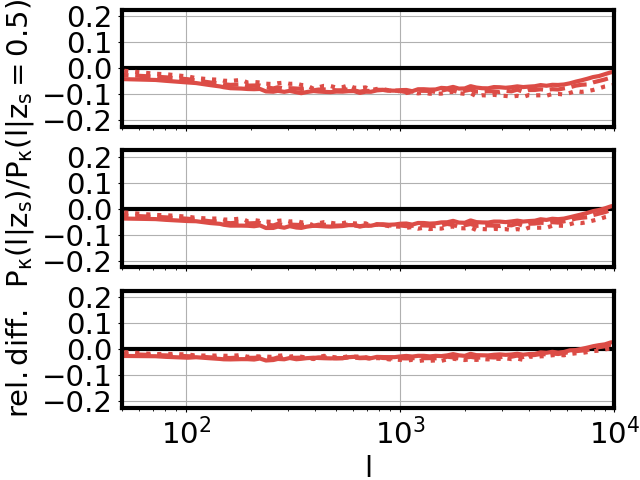}
  \includegraphics[width=0.49\hsize]{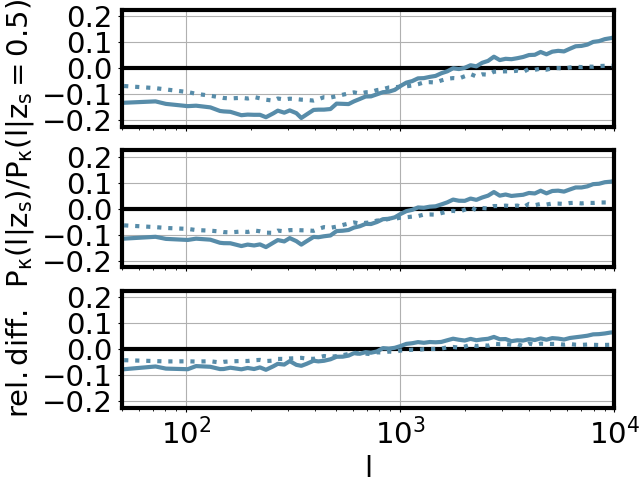}
  \caption{Relative difference between  the convergence power spectrum
    at  a   given  source  redshift  and   that  measured  considering
    $z_s=0.5$. Top left panel show the pure MG models, while the other
    panels display the cosmologies combining MG and massive neutrinos,
    colour-coded in green, red and blue for the $fR6$, $fR5$ and $fR4$
    model, respectively.\label{Cfigpkl}}
\end{figure*}

In  Fig.~\ref{Dfigpkl}  we  show   the  relative  differences  of  the
convergence power  spectra as computed for  the different cosmological
models with respect to the standard  \lcdm one.  In the top-left panel
we  display the  relative  difference  of the  pure  MG models;  $z_s$
decreases going  from top to  bottom. In the sub-panels  for $z_s=0.5$
and $z_s=1$  we display also  the shape measurement  uncertainties for
$8$  and $33$  galaxies per  square arcmin,  shaded in  cyan and  pink
respectively,  up  to $l=3\times  10^3$  that  represents the  highest
angular mode expected to be probed by a survey like Euclid.  From this
figure we  notice that, while  for a low-redshift weak  lensing survey
with  a limited  source  number  density it  is  quite challenging  to
discriminate between the various MG  models, a wide space-based survey
with much larger number density of sources for shape measurements will
be certainly able to distinguish $fR4$ and $fR5$, but still likely not
$fR6$. The other  three panels display the relative  difference of all
the other  models with respect  to the \lcdm  one, for the  same fixed
source redshifts. In green, red and  blue we show the $fR6$, $fR5$ and
$fR4$ models with  and without massive neutrinos. From  the figures we
can also  notice that  it will  be very difficult  for a  future space
survey  to discriminate  the  $fR6$ models  with  and without  massive
neutrinos,  as   well  as  the  $fR4\_0.3$eV   and  the  $fR5\_0.15$eV
\citep[see  also  our  companion paper][]{peel18}.   This  is  another
example  of how  the  degeneracy between  $f(R)$  gravity and  massive
neutrinos can  affect the  constraining power  of some  observables in
future  cosmological observations,  and  shows the  need of  combining
primary cosmological  probes with  other observables like  the cluster
counts --  as discussed  in Section~\ref{halocatalogues};  in addition
these  observed degeneracies  strongly motivate  the need  of devising
more  sophisticated statistics  of the  WL signals  that may  allow to
break them.   In this context it  will be necessary to  look at higher
order statistics,  probing the  shape of the  probability distribution
function  of  the convergence  that  are  found  to provide  a  higher
discriminating  power for  the $f(R)-m_{\nu  }$ degeneracy  \citep[see
  e.g.][]{peel18}, besides  their known  feature to break  the $\Omega
_{\rm   M}-\sigma   _{8}$    degeneracy   within   \lcdm   cosmologies
\citep{giocoli18a,vicinanza18}.  Alternatively, one may try to exploit
a different leverage, namely the possibly different redshift evolution
of  the  $f(R)$  and  massive neutrino  signals,  to  disentangle  the
degenerate models from \lcdm.

From the power spectra at different  source redshifts, in fact, we can
see that the  relative difference with respect to \lcdm  changes, as a
consequence  of the  redshift evolution  of MG  and massive  neutrinos
effects.   In the  bottom right  panel  we see  that the  peak in  the
relative difference of  the projected power spectra of  $fR4$ moves to
smaller scales (larger $l$ modes) when the source redshift increases.

The  relative  difference of  the  convergence  power spectra  of  the
various  models at  fixed angular  mode as  a function  of the  source
redshift is  displayed in Fig.~\ref{figlfixed}. Left  and right panels
show the relative difference of the various models with respect to the
\lcdm one  at $l=100$ and  $l=1000$, respectively.  Almost  all models
display  a monotonic  trend, decreasing  as a  function of  the source
redshifts. The relative  difference of the MG  models (without massive
neutrinos) tends to be larger at  smaller scales -- larger $l$, and it
is  reduced in  magnitude  by  the presence  of  the massive  neutrino
components  depending on  the neutrino  mass value.   As expected,  we
notice  that the  $fR6$ model  presents the  smallest difference  with
respect to the \lcdm one as a  function of $z_s$. The right panel also
shows that while for high  redshift sources $fR6$ is indistinguishable
from  \lcdm,  for  $z_s$  moving toward  low  redshifts  the  relative
difference can be as large as approximately $5\%$.

In Fig.~\ref{Cfigpkl} we display the  relative difference of the ratio
between the convergence power spectrum  at a given source redshift and
the power spectrum measured at  low redshift, i.e.  $z_s=0.5$, for the
various cosmological models. Therefore,  these plots show the relative
growth of the projected matter density distribution with redshift with
respect to the standard cosmology. As  in the previous figure, the top
left sub-panels show  the relative measurements of the  pure MG alone,
while in  the other  panels display  for each  colour the  models that
account for  both MG and  massive neutrino components.  We  can notice
that the relative  trends are redshift dependent,  the difference with
respect  to  \lcdm  is  larger   at  higher  source  redshifts.   Most
importantly,  we can  notice that  this kind  of tomographic  approach
allows  to break  the degeneracy  between $f(R)$  gravity and  massive
neutrinos, as it  can be seen by  the fact that the  introduction of a
massive neutrino component does not  make the relative difference with
respect to the  \lcdm case to vanish at all  scales, in particular for
the $fR4$ and  $fR5$ models, where a clear signature  of the MG effect
remains detectable.

\begin{figure*}
\centering
\hspace{-0.4cm}
  \includegraphics[width=0.315\hsize]{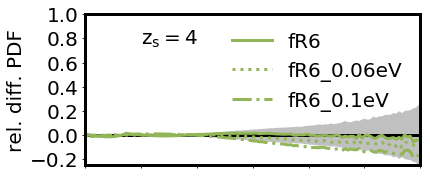}
  \hspace{0.17cm}  
  \includegraphics[width=0.315\hsize]{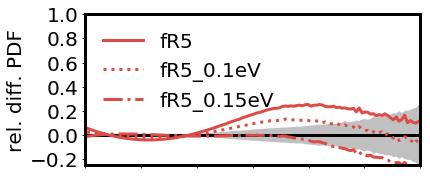}
  \hspace{0.14cm}
  \includegraphics[width=0.315\hsize]{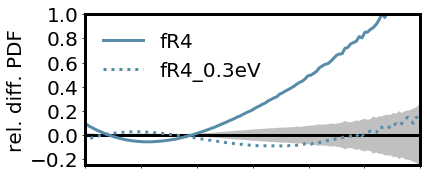}\\
  \hspace{-0.4cm}
  \includegraphics[width=0.315\hsize]{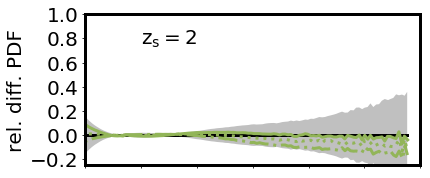}  
   \hspace{0.17cm}
  \includegraphics[width=0.315\hsize]{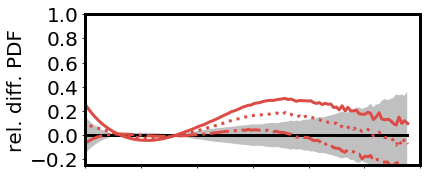}
  \hspace{0.14cm}
  \includegraphics[width=0.315\hsize]{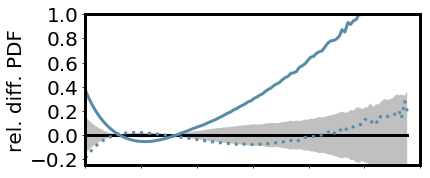}\\
  \includegraphics[width=0.328\hsize]{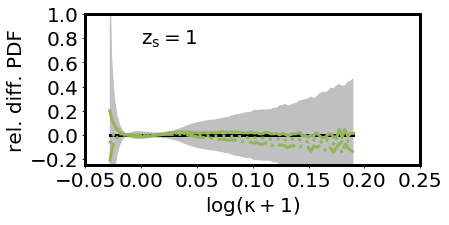}
  \includegraphics[width=0.328\hsize]{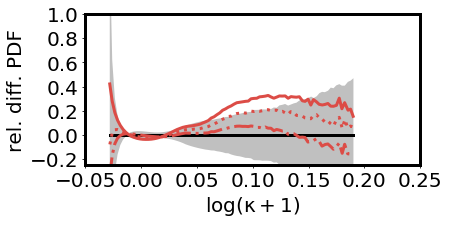}
  \includegraphics[width=0.328\hsize]{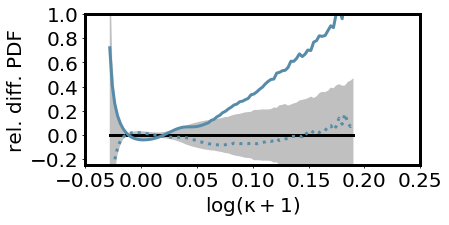}
  \caption{Relative difference of the convergence distribution for the
    different cosmological models  in term of the \lcdm  one. From top
    to bottom we display the prediction  for $z_s=4$, $2$ and $1$. The
    curves  refer  to  the   average  distribution  over  the  various
    realisations while the shaded area marks the variance of the \lcdm
    measurements.  The  data are constructed  binning $\log(\kappa+1)$
    between $-0.15$  and $0.5$ in  $256$ intervals and  displayed when
    for the \lcdm  model the counts are larger than  $10$ units. Green
    (left panels), blue (central panels)  and red (right panels) lines
    refer to the various MG models, while the corresponding dotted and
    dashed   curves    display   the   results   for    the   combined
    cosmologies. \label{figpdf}}
\end{figure*}

In  Fig.~\ref{figpdf}  we  show   the  PDF  (Probability  Distribution
Function)  of   the  relative  difference  of   the  convergence  maps
constructed at three  source redshift $z_s=1$, $2$ and  $4$.  To avoid
overcrowding  the figure  we  decided  not to  display  the cases  for
$z_s=0.5$, that  do not  show particular  differences with  respect to
\lcdm,  apart  for the  extreme  modified  gravity model  $fR4$.   The
histograms are  computed by binning the  quantity $\log(\kappa+1)$ for
each pixel between  $-0.15$ and $0.5$ in $256$  intervals.  The panels
on  the  left display  the  $fR6$  models,  with and  without  massive
neutrinos, from  which we  can notice  that their  relative difference
with respect to  \lcdm is again very well within  the sample variance,
represented by the shaded gray  regions. The middle panels display the
measurements for the $fR5$ cosmologies, where we can notice a bump for
$0.25<\kappa <0.42$, that decreases with  the source redshift and with
the total neutrino  mass. This may be an important  signature that can
be detectable in the magnification distributions and image parities of
strong lensing, as discussed by  \citet{castro17}.  In the right panel
we display  the relative trend of  the PDF of the  convergence for the
$fR4$  models. The  cosmology without  massive neutrino  component has
much   more  compact   non-linear  systems   and  a   difference  that
monotonically increases with the convergence from $\kappa \approx0.12$
reaching  values  of  $100\%$  already  at  $\kappa\approx0.58$.   The
one-point  statistic of  the  convergence field  for the  $fR4\_0.3eV$
model shows again a strong degeneracy -- within the sample variance --
with respect to  \lcdm at low redshifts, while for  $z_s=4$ it shows a
relative difference  reaching $5-10\%$, positive for  $k\approx 0$ and
negative  for  $k\approx  0.25$.   In this  final  paragraph  we  have
discussed  the PDF  distributions of  the convergence  -- for  various
source  redshifts;  these  are  simpler   to  work  with  and  provide
complementary information  to the convergence power  spectra affecting
for instance the incidence of multiple image events.  The cosmological
calibration  and  characterization  of   the  lensing  PDFs  represent
important studies for future wide fields  surveys: they can be used to
infer  valuable  information  on  the large-scale  structure  and  its
evolution  through  its  effect  on standard  candles  (like  type  Ia
Supernovae)  since it  introduces non-Gaussianities  to their  scatter
\citep{bernardeau97,hamana00,valageas00}.

\section{Summary and Conclusions}
\label{final}

In this  paper we have  presented a  new suite of  cosmological N-body
simulations -- the \dustp runs -- and we have discussed a first set of
results from  weak lensing  and halo catalogs  within past-light-cones
extracted from  their outputs.   The {\small  DUSTGRAIN} project  is a
numerical   enterprise  aimed   at   the   investigation  of   various
cosmological  observables  in  the   context  of  cosmological  models
featuring at  the same time a  non-standard theory of gravity  -- here
assumed to  be in the  form of \citeauthor{Hu_Sawicki_2007}  $f(R)$ --
and  a non-vanishing  neutrino mass,  with the  final goal  of testing
observational degeneracies between these two independent phenomena. In
particular, the {\em pathfinder} runs are a suite of intermediate-size
simulations  aimed at  sampling  the joint  parameter  space of  these
combined models in order to select the most degenerate combinations of
their  characteristic  parameters $f_{R0}$  and  $m_{nu}$  to be  then
resimulated woth higher dynamical range in the {\small DUSTGRAIN} full
scale runs. Nonetheless, these preliminary simulations already deliver
a wealth of novel information to deserve a detailed analysis.

In particular, in  this work we have focussed on  halo counts and weak
lensing statistics within the past  light cones. These extend from the
observer -- located  at $z=0$ -- up to redshift  $z=4$ with an angular
aperture of $5\times  5$ sq.  degree resolved with  $2048 \times 2048$
pixels. From each simulation we  generated a sample of $256$ different
realisations  randomising   the  boxes   of  the   various  simulation
snapshots.

In what follows we summarize our main results:
\begin{itemize}

\item  the   three-dimensional  comoving  power  spectra   show  clear
  signatures  of Modified  Gravity  with a  difference that  increases
  towards larger scales  and lower redshifts; the  presence of massive
  neutrinos,  however,  slows down  the  growth  rate of  perturbation
  reducing the matter power spectrum  mainly at large scales; strongly
  suppresses  these observational  footprints;  therefore, the  matter
  power spectra show a strong  degeneracy with respect to the standard
  \lcdm scenario for suitable combinations of MG and massive neutrinos
  parameters, as already found by previous studies;
  
\item   the  2D   projected   halo  mass   function  shows   different
  characteristic signatures for different  modified gravity models. In
  particular, the $fR6$  model shows an increase of  the low-mass halo
  counts while  the $fR5$  model presents at  low (high)  redshifts an
  excess      of      $\approx      10^{14}M_{\odot}/h$      ($\approx
  10^{13}M_{\odot}/h$) haloes  and the $fR4$ run  displays a monotonic
  increase of the halo mass function in the light-cone with respect to
  the standard  \lcdm with  a clear  excess of  high mass  haloes; the
  inclusion  of  massive neutrinos  always  suppresses  the counts  of
  high-mass halos resulting in a  closer agreement with respect to the
  reference \lcdm model, except for $fR6$;

  \item  the redshift  distribution  of galaxy  clusters (haloes  with
    $M_{200}>10^{14}M_{\odot}/h$) in the $fR4$ model is very different
    from \lcdm with an excess  of approximately $40\%$ with respect to
    the reference model; all the  other cosmologies present a redshift
    distribution of galaxy clusters within  the sample variance of the
    \lcdm one apart  from $fR5$ with $m_{\nu }0.15$ eV  and $fR4$ with
    $m_{\nu }=  0.3$ eV that display  a decline of counts  toward high
    redshifts;

  \item  the convergence  power  spectra show  distinct signatures  of
    Modified  Gravity, mainly  for  high source  redshifts, where  the
    different   non-linear   properties    of   the   structures   are
    distinguishable for angular modes $l\ge  10^3$; also in this case,
    the  presence of  massive neutrinos  suppresses these  signatures,
    making the measurements closer to the standard reference model;

  \item  a tomographic  analysis  of the  power  spectra at  different
    source redshifts is  found to be very  promising in distinguishing
    some  of  these  degenerate  non-standard  models  from  \lcdm  by
    exploiting the different redshift  evolution of the $f(R)$ gravity
    and massive neutrinos effects  on large-scale structures formation
    and evolution; the  same is also true for  the one-point statistic
    of the convergence field for different source redshifts.
   
\end{itemize}

The analysis of simulated past-light-cones of non-standard cosmologies
represents a realistic  way to assess the capabilities  of ongoing and
future surveys  to disentangle such cosmological  models from standard
\lcdm. In this  work we have analysed cluster counts  and weak lensing
statistics of $f(R)$ gravity  cosmological N-body simulations with and
without massive  neutrinos and compared  the findings with  a standard
reference \lcdm  cosmology. In  both cluster  counts and  weak lensing
statistics some  of the pure  $f(R)$ models appear very  distinct from
\lcdm, but  the combination  with massive neutrinos  can make  them to
appear  again consistent  with  the standard  scenario. These  results
underline the  difficulty of  the 'observables' we  have looked  at to
disentangle  by   themselves  these  degenerate   non-standard  models
suggesting the  need of combining  various cosmological probes,  or of
looking  at higher  order  statistics or  cross correlating  different
observables, or finally to exploit observations ad different redshifts
to break the degeneracies.

\section*{Acknowledgements}                                           
CG and MB acknowledge support from the Italian Ministry for Education,
University  and  Research  (MIUR)  through the  SIR  individual  grant
SIMCODE (project number RBSI14P4IH).  CG acknowledges support from the
Italian  Ministry of  Foreign Affairs  and International  Cooperation,
Directorate General  for Country  Promotion.  We also  acknowledge the
support  from the  grant  MIUR PRIN  2015  "Cosmology and  Fundamental
Physics:  illuminating  the  Dark   Universe  with  Euclid";  and  the
financial contribution from the agreement ASI n.I/023/12/0 "Attivit\`a
relative  alla  fase  B2/C  per   la  missione  Euclid".   The  \dustp
simulations discussed in this work have been performed and analysed on
the  Marconi supercomputing  machine  at Cineca  thanks  to the  PRACE
project  SIMCODE1  (grant  nr.    2016153604)  and  on  the  computing
facilities of  the Computational Center for  Particle and Astrophysics
(C2PAP)  and  of the  Leibniz  Supercomputer  Center (LRZ)  under  the
project ID pr94ji.  We thank the anonymous reviewer for her/his useful
comments.

\bibliographystyle{mn2e}

\bsp	
\bibliography{plcpathfinder}

\begin{thebibliography}{}

\bibitem[\protect\citeauthoryear{{Arnold}, {Fosalba}, {Springel}, {Puchwein} \&
  {Blot}}{{Arnold} et~al.}{2018}]{Arnold_etal_2018}
{Arnold} C.,  {Fosalba} P.,  {Springel} V.,  {Puchwein} E.,    {Blot} L.,
  2018, ArXiv e-prints 1805.09824

\bibitem[\protect\citeauthoryear{{Arnold}, {Puchwein} \& {Springel}}{{Arnold}
  et~al.}{2014}]{Arnold_Puchwein_Springel_2014}
{Arnold} C.,  {Puchwein} E.,    {Springel} V.,  2014, \mnras, 440, 833

\bibitem[\protect\citeauthoryear{Arnold, Puchwein \& Springel}{Arnold
  et~al.}{2015}]{Arnold_Puchwein_Springel_2015}
Arnold C.,  Puchwein E.,    Springel V.,  2015, Mon. Not. Roy. Astron. Soc.,
  448, 2275

\bibitem[\protect\citeauthoryear{{Arnold}, {Springel} \& {Puchwein}}{{Arnold}
  et~al.}{2016}]{Arnold_Springel_Puchwein_2016}
{Arnold} C.,  {Springel} V.,    {Puchwein} E.,  2016, \mnras, 462, 1530

\bibitem[\protect\citeauthoryear{{Baldi} \& {Villaescusa-Navarro}}{{Baldi} \&
  {Villaescusa-Navarro}}{2018}]{Baldi_Villaescusa-Navarro_2018}
{Baldi} M.,  {Villaescusa-Navarro} F.,  2018, \mnras, 473, 3226

\bibitem[\protect\citeauthoryear{Baldi, Villaescusa-Navarro, Viel, Puchwein,
  Springel \& Moscardini}{Baldi et~al.}{2014}]{Baldi_etal_2014}
Baldi M.,  Villaescusa-Navarro F.,  Viel M.,  Puchwein E.,  Springel V.,
  Moscardini L.,  2014, Mon. Not. Roy. Astron. Soc., 440, 75

\bibitem[\protect\citeauthoryear{{Barreira}, {Bose}, {Li} \&
  {Llinares}}{{Barreira} et~al.}{2017}]{barreira17}
{Barreira} A.,  {Bose} S.,  {Li} B.,    {Llinares} C.,  2017, \jcap, 2, 031

\bibitem[\protect\citeauthoryear{{Barreira}, {Llinares}, {Bose} \&
  {Li}}{{Barreira} et~al.}{2016}]{Barreira_etal_2016}
{Barreira} A.,  {Llinares} C.,  {Bose} S.,    {Li} B.,  2016, \jcap, 5, 001

\bibitem[\protect\citeauthoryear{{Bartelmann}}{{Bartelmann}}{2010}]{bartelmann10}
{Bartelmann} M.,  2010, Classical and Quantum Gravity, 27, 233001

\bibitem[\protect\citeauthoryear{{Bartelmann} \& {Schneider}}{{Bartelmann} \&
  {Schneider}}{2001}]{bartelmann01}
{Bartelmann} M.,  {Schneider} P.,  2001, Physics Reports, 340, 291

\bibitem[\protect\citeauthoryear{{Bellomo}, {Bellini}, {Hu}, {Jimenez},
  {Pena-Garay} \& {Verde}}{{Bellomo} et~al.}{2017}]{Bellomo_etal_2017}
{Bellomo} N.,  {Bellini} E.,  {Hu} B.,  {Jimenez} R.,  {Pena-Garay} C.,
  {Verde} L.,  2017, \jcap, 2, 043

\bibitem[\protect\citeauthoryear{{Benjamin}, {Van Waerbeke}, {Heymans},
  {Kilbinger}, {Erben}, {Hildebrandt}, {Hoekstra} \& {et al.}}{{Benjamin}
  et~al.}{2013}]{benjamin13}
{Benjamin} J.,  {Van Waerbeke} L.,  {Heymans} C.,  {Kilbinger} M.,  {Erben} T.,
   {Hildebrandt} H.,  {Hoekstra} H.,    {et al.} 2013, \mnras, 431, 1547

\bibitem[\protect\citeauthoryear{Bernardeau, Van~Waerbeke \&
  Mellier}{Bernardeau et~al.}{1997}]{bernardeau97}
Bernardeau F.,  Van~Waerbeke L.,    Mellier Y.,  1997, Astron.Astrophys., 322,
  1

\bibitem[\protect\citeauthoryear{Bertotti, Iess \& Tortora}{Bertotti
  et~al.}{2003}]{Bertotti_Iess_Tortora_2003}
Bertotti B.,  Iess L.,    Tortora P.,  2003, Nature, 425, 374

\bibitem[\protect\citeauthoryear{{Bertschinger} \& {Zukin}}{{Bertschinger} \&
  {Zukin}}{2008}]{bertschinger08}
{Bertschinger} E.,  {Zukin} P.,  2008, \prd, 78, 024015

\bibitem[\protect\citeauthoryear{{Brax} \& {Valageas}}{{Brax} \&
  {Valageas}}{2014}]{Brax_Valageas_2014}
{Brax} P.,  {Valageas} P.,  2014, \prd, 90, 023508

\bibitem[\protect\citeauthoryear{{Broadhurst}, {Ben{\'{\i}}tez}, {Coe},
  {Sharon}, {Zekser}, {White}, {Ford} \& et al.}{{Broadhurst}
  et~al.}{2005b}]{broadhurst05b}
{Broadhurst} T.,  {Ben{\'{\i}}tez} N.,  {Coe} D.,  {Sharon} K.,  {Zekser} K.,
  {White} R.,  {Ford} H.,    et al. B.,  2005b, \apj, 621, 53

\bibitem[\protect\citeauthoryear{{Buchdahl}}{{Buchdahl}}{1970}]{Buchdahl_1970}
{Buchdahl} H.~A.,  1970, \mnras, 150, 1

\bibitem[\protect\citeauthoryear{{Castro}, {Quartin}, {Giocoli}, {Borgani} \&
  {Dolag}}{{Castro} et~al.}{2018}]{castro17}
{Castro} T.,  {Quartin} M.,  {Giocoli} C.,  {Borgani} S.,    {Dolag} K.,  2018,
  \mnras, 478, 1305

\bibitem[\protect\citeauthoryear{{Cleveland}, {Daily}, {Davis} Jr., {Distel},
  {Lande}, {Lee}, {Wildenhain} \& {Ullman}}{{Cleveland}
  et~al.}{1998}]{Cleveland_etal_1998}
{Cleveland} B.~T.,  {Daily} T.,  {Davis} Jr. R.,  {Distel} J.~R.,  {Lande} K.,
  {Lee} C.~K.,  {Wildenhain} P.~S.,    {Ullman} J.,  1998, \apj, 496, 505

\bibitem[\protect\citeauthoryear{{Coe}, {Zitrin}, {Carrasco}, {Shu}, {Zheng},
  {Postman}, {Bradley}, {Koekemoer} \& et al.}{{Coe} et~al.}{2013}]{coe13}
{Coe} D.,  {Zitrin} A.,  {Carrasco} M.,  {Shu} X.,  {Zheng} W.,  {Postman} M.,
  {Bradley} L.,  {Koekemoer}   et al. 2013, \apj, 762, 32

\bibitem[\protect\citeauthoryear{{Costanzi}, {Sartoris}, {Viel} \&
  {Borgani}}{{Costanzi} et~al.}{2014}]{costanzi14}
{Costanzi} M.,  {Sartoris} B.,  {Viel} M.,    {Borgani} S.,  2014, \jcap, 10,
  081

\bibitem[\protect\citeauthoryear{{Davis}, {Efstathiou}, {Frenk} \&
  White}{{Davis} et~al.}{1985}]{Davis_etal_1985}
{Davis} M.,  {Efstathiou} G.,  {Frenk} C.~S.,    White S.~D.,  1985,
  Astrophys.J., 292, 371

\bibitem[\protect\citeauthoryear{Deffayet, Dvali, Gabadadze \&
  Vainshtein}{Deffayet et~al.}{2002}]{Deffayet_etal_2002}
Deffayet C.,  Dvali G.,  Gabadadze G.,    Vainshtein A.~I.,  2002, Phys.Rev.,
  D65, 044026

\bibitem[\protect\citeauthoryear{{Despali}, {Giocoli}, {Angulo}, {Tormen},
  {Sheth}, {Baso} \& {Moscardini}}{{Despali} et~al.}{2016}]{despali16}
{Despali} G.,  {Giocoli} C.,  {Angulo} R.~E.,  {Tormen} G.,  {Sheth} R.~K.,
  {Baso} G.,    {Moscardini} L.,  2016, \mnras, 456, 2486

\bibitem[\protect\citeauthoryear{{Einstein}}{{Einstein}}{1918}]{einstein18}
{Einstein} A.,  1918, Sitzungsberichte der K{\"o}niglich Preu{\ss}ischen
  Akademie der Wissenschaften (Berlin), Seite 448-459.

\bibitem[\protect\citeauthoryear{{Erben}, {Hildebrandt}, {Miller}, {van
  Waerbeke}, {Heymans}, {Hoekstra}, {Kitching} \& {et al.}}{{Erben}
  et~al.}{2013}]{erben12}
{Erben} T.,  {Hildebrandt} H.,  {Miller} L.,  {van Waerbeke} L.,  {Heymans} C.,
   {Hoekstra} H.,  {Kitching} T.~D.,    {et al.} 2013, \mnras, 433, 2545

\bibitem[\protect\citeauthoryear{{Fu}, {Semboloni}, {Hoekstra}, {Kilbinger},
  {van Waerbeke}, {Tereno}, {Mellier}, {Heymans}, {Coupon}, {Benabed},
  {Benjamin}, {Bertin}, {Dor{\'e}}, {Hudson}, {Ilbert}, {Maoli} \&
  {et~al.}}{{Fu} et~al.}{2008}]{fu08}
{Fu} L.,  {Semboloni} E.,  {Hoekstra} H.,  {Kilbinger} M.,  {van Waerbeke} L.,
  {Tereno} I.,  {Mellier} Y.,  {Heymans} C.,  {Coupon} J.,  {Benabed} K.,
  {Benjamin} J.,  {Bertin} E.,  {Dor{\'e}} O.,  {Hudson} M.~J.,  {Ilbert} O.,
  {Maoli}   {et~al.} 2008, \aap, 479, 9

\bibitem[\protect\citeauthoryear{{Giocoli}, {Di Meo}, {Meneghetti}, {Jullo},
  {de la Torre}, {Moscardini}, {Baldi}, {Mazzotta} \& {Metcalf}}{{Giocoli}
  et~al.}{2017}]{giocoli17}
{Giocoli} C.,  {Di Meo} S.,  {Meneghetti} M.,  {Jullo} E.,  {de la Torre} S.,
  {Moscardini} L.,  {Baldi} M.,  {Mazzotta} P.,    {Metcalf} R.~B.,  2017,
  \mnras, 470, 3574

\bibitem[\protect\citeauthoryear{{Giocoli}, {Jullo}, {Metcalf}, {de la Torre},
  {Yepes}, {Prada}, {Comparat}, {G{\"o}ttlober}, {Kyplin}, {Kneib}, {Petkova},
  {Shan} \& {Tessore}}{{Giocoli} et~al.}{2016}]{giocoli16a}
{Giocoli} C.,  {Jullo} E.,  {Metcalf} R.~B.,  {de la Torre} S.,  {Yepes} G.,
  {Prada} F.,  {Comparat} J.,  {G{\"o}ttlober} S.,  {Kyplin} A.,  {Kneib}
  J.-P.,  {Petkova} M.,  {Shan} H.~Y.,    {Tessore} N.,  2016, \mnras, 461, 209

\bibitem[\protect\citeauthoryear{{Giocoli}, {Meneghetti}, {Metcalf}, {Ettori}
  \& {Moscardini}}{{Giocoli} et~al.}{2014}]{giocoli14}
{Giocoli} C.,  {Meneghetti} M.,  {Metcalf} R.~B.,  {Ettori} S.,    {Moscardini}
  L.,  2014, \mnras, 440, 1899

\bibitem[\protect\citeauthoryear{{Giocoli}, {Moreno}, {Sheth} \&
  {Tormen}}{{Giocoli} et~al.}{2007}]{giocoli07}
{Giocoli} C.,  {Moreno} J.,  {Sheth} R.~K.,    {Tormen} G.,  2007, \mnras, 376,
  977

\bibitem[\protect\citeauthoryear{{Giocoli}, {Moscardini}, {Baldi}, {Meneghetti}
  \& {Metcalf}}{{Giocoli} et~al.}{2018}]{giocoli18a}
{Giocoli} C.,  {Moscardini} L.,  {Baldi} M.,  {Meneghetti} M.,    {Metcalf}
  R.~B.,  2018, \mnras

\bibitem[\protect\citeauthoryear{{Hagstotz}, {Costanzi}, {Baldi} \&
  {Weller}}{{Hagstotz} et~al.}{2018}]{hagstotz18}
{Hagstotz} S.,  {Costanzi} M.,  {Baldi} M.,    {Weller} J.,  2018, eprints
  ArXiv: 1806.07400

\bibitem[\protect\citeauthoryear{Hamana \& Futamase}{Hamana \&
  Futamase}{2000}]{hamana00}
Hamana T.,  Futamase T.,  2000, ApJ, 534, 29

\bibitem[\protect\citeauthoryear{{Harnois-D{\'e}raps}, {Munshi}, {Valageas},
  {van Waerbeke}, {Brax}, {Coles} \& {Rizzo}}{{Harnois-D{\'e}raps}
  et~al.}{2015}]{harnois-deraps15c}
{Harnois-D{\'e}raps} J.,  {Munshi} D.,  {Valageas} P.,  {van Waerbeke} L.,
  {Brax} P.,  {Coles} P.,    {Rizzo} L.,  2015, \mnras, 454, 2722

\bibitem[\protect\citeauthoryear{{He}}{{He}}{2013}]{He_2013}
{He} J.-h.,  2013, arXiv:1307.4876

\bibitem[\protect\citeauthoryear{{Heymans}, {Grocutt}, {Heavens}, {Kilbinger},
  {Kitching}, {Simpson}, {Benjamin}, {Erben} \& {et al.}}{{Heymans}
  et~al.}{2013}]{heymans13}
{Heymans} C.,  {Grocutt} E.,  {Heavens} A.,  {Kilbinger} M.,  {Kitching} T.~D.,
   {Simpson} F.,  {Benjamin} J.,  {Erben} T.,    {et al.} 2013, \mnras, 432,
  2433

\bibitem[\protect\citeauthoryear{{Higuchi} \& {Shirasaki}}{{Higuchi} \&
  {Shirasaki}}{2016}]{higuchi16}
{Higuchi} Y.,  {Shirasaki} M.,  2016, \mnras, 459, 2762

\bibitem[\protect\citeauthoryear{{Hildebrandt}, {Viola}, {Heymans}, {Joudaki},
  {Kuijken}, {Blake}, {Erben}, {Joachimi} \& {et al.}}{{Hildebrandt}
  et~al.}{2017}]{hildebrandt17}
{Hildebrandt} H.,  {Viola} M.,  {Heymans} C.,  {Joudaki} S.,  {Kuijken} K.,
  {Blake} C.,  {Erben} T.,  {Joachimi} B.,    {et al.} 2017, \mnras, 465, 1454

\bibitem[\protect\citeauthoryear{Hinterbichler \& Khoury}{Hinterbichler \&
  Khoury}{2010}]{Hinterbichler_Khoury_2010}
Hinterbichler K.,  Khoury J.,  2010, Phys.Rev.Lett., 104, 231301

\bibitem[\protect\citeauthoryear{{Hojjati}, {Pogosian} \& {Zhao}}{{Hojjati}
  et~al.}{2011}]{hojjati11}
{Hojjati} A.,  {Pogosian} L.,    {Zhao} G.-B.,  2011, \jcap, 8, 005

\bibitem[\protect\citeauthoryear{Hu \& Sawicki}{Hu \&
  Sawicki}{2007}]{Hu_Sawicki_2007}
Hu W.,  Sawicki I.,  2007, Phys. Rev., D76, 064004

\bibitem[\protect\citeauthoryear{{Huterer}}{{Huterer}}{2002}]{huterer02}
{Huterer} D.,  2002, \prd, 65, 063001

\bibitem[\protect\citeauthoryear{{Kaiser}, {Squires} \& {Broadhurst}}{{Kaiser}
  et~al.}{1995}]{kaiser95}
{Kaiser} N.,  {Squires} G.,    {Broadhurst} T.,  1995, \apj, 449, 460

\bibitem[\protect\citeauthoryear{Khoury \& Weltman}{Khoury \&
  Weltman}{2004}]{Khoury_Weltman_2004}
Khoury J.,  Weltman A.,  2004, Phys.Rev., D69, 044026

\bibitem[\protect\citeauthoryear{{Kilbinger}}{{Kilbinger}}{2015}]{kilbinger14}
{Kilbinger} M.,  2015, Reports on Progress in Physics, 78, 086901

\bibitem[\protect\citeauthoryear{{Kilbinger}, {Fu}, {Heymans}, {Simpson},
  {Benjamin}, {Erben}, {Harnois-D{\'e}raps}, {Hoekstra}, {Hildebrandt} \&
  {et~al.}}{{Kilbinger} et~al.}{2013}]{kilbinger13}
{Kilbinger} M.,  {Fu} L.,  {Heymans} C.,  {Simpson} F.,  {Benjamin} J.,
  {Erben} T.,  {Harnois-D{\'e}raps} J.,  {Hoekstra} H.,  {Hildebrandt} H.,
  {et~al.} 2013, \mnras, 430, 2200

\bibitem[\protect\citeauthoryear{{Kitching}, {Heavens}, {Alsing}, {Erben},
  {Heymans}, {Hildebrandt}, {Hoekstra} \& {et al.}}{{Kitching}
  et~al.}{2014}]{kitching14}
{Kitching} T.~D.,  {Heavens} A.~F.,  {Alsing} J.,  {Erben} T.,  {Heymans} C.,
  {Hildebrandt} H.,  {Hoekstra} H.,    {et al.} 2014, \mnras, 442, 1326

\bibitem[\protect\citeauthoryear{{Kitching}, {Taylor}, {Cropper}, {Hoekstra},
  {Hood}, {Massey} \& {Niemi}}{{Kitching} et~al.}{2016}]{kitching16}
{Kitching} T.~D.,  {Taylor} A.~N.,  {Cropper} M.,  {Hoekstra} H.,  {Hood}
  R.~K.~E.,  {Massey} R.,    {Niemi} S.,  2016, \mnras, 455, 3319

\bibitem[\protect\citeauthoryear{{Landau} \& {Lifshitz}}{{Landau} \&
  {Lifshitz}}{1971}]{landau71}
{Landau} L.~D.,  {Lifshitz} E.~M.,  1971, {The classical theory of fields}

\bibitem[\protect\citeauthoryear{{Laureijs}, {Amiaux}, {Arduini},
  {Augu{\`e}res}, {Brinchmann}, {Cole}, {Cropper}, {Dabin}, {Duvet} \& et
  al.}{{Laureijs} et~al.}{2011}]{euclidredbook}
{Laureijs} R.,  {Amiaux} J.,  {Arduini} S.,  {Augu{\`e}res} J.~.,  {Brinchmann}
  J.,  {Cole} R.,  {Cropper} M.,  {Dabin} C.,  {Duvet} L.,    et al. 2011,
  eprint arXiv: 1110.3193

\bibitem[\protect\citeauthoryear{{Lesgourgues} \& {Pastor}}{{Lesgourgues} \&
  {Pastor}}{2006}]{lesgourgues06}
{Lesgourgues} J.,  {Pastor} S.,  2006, \physrep, 429, 307

\bibitem[\protect\citeauthoryear{Lewis, Challinor \& Lasenby}{Lewis
  et~al.}{2000}]{camb}
Lewis A.,  Challinor A.,    Lasenby A.,  2000, Astrophys. J., 538, 473

\bibitem[\protect\citeauthoryear{{Li} \& {Shirasaki}}{{Li} \&
  {Shirasaki}}{2018}]{li18}
{Li} B.,  {Shirasaki} M.,  2018, \mnras, 474, 3599

\bibitem[\protect\citeauthoryear{Li, Zhao, Teyssier \& Koyama}{Li
  et~al.}{2012}]{Ecosmog}
Li B.,  Zhao G.-B.,  Teyssier R.,    Koyama K.,  2012, JCAP, 1201, 051

\bibitem[\protect\citeauthoryear{{Lin} \& {Kilbinger}}{{Lin} \&
  {Kilbinger}}{2015}]{lin15a}
{Lin} C.-A.,  {Kilbinger} M.,  2015, \aap, 576, A24

\bibitem[\protect\citeauthoryear{Llinares, Mota \& Winther}{Llinares
  et~al.}{2014}]{Llinares_Mota_Winther_2014}
Llinares C.,  Mota D.~F.,    Winther H.~A.,  2014, Astron. Astrophys., 562, A78

\bibitem[\protect\citeauthoryear{{Mead}, {Heymans}, {Lombriser}, {Peacock},
  {Steele} \& {Winther}}{{Mead} et~al.}{2016}]{mead16}
{Mead} A.~J.,  {Heymans} C.,  {Lombriser} L.,  {Peacock} J.~A.,  {Steele}
  O.~I.,    {Winther} H.~A.,  2016, \mnras, 459, 1468

\bibitem[\protect\citeauthoryear{{Mead}, {Peacock}, {Heymans}, {Joudaki} \&
  {Heavens}}{{Mead} et~al.}{2015}]{mead15b}
{Mead} A.~J.,  {Peacock} J.~A.,  {Heymans} C.,  {Joudaki} S.,    {Heavens}
  A.~F.,  2015, \mnras, 454, 1958

\bibitem[\protect\citeauthoryear{{Meneghetti}, {Bartelmann}, {Dahle} \&
  {Limousin}}{{Meneghetti} et~al.}{2013}]{meneghetti13}
{Meneghetti} M.,  {Bartelmann} M.,  {Dahle} H.,    {Limousin} M.,  2013, \ssr:
  ArXiv e-print: 1303.3363

\bibitem[\protect\citeauthoryear{Motohashi, Starobinsky \& Yokoyama}{Motohashi
  et~al.}{2013}]{Motohashi_etal_2013}
Motohashi H.,  Starobinsky A.~A.,    Yokoyama J.,  2013, Phys.Rev.Lett., 110,
  121302

\bibitem[\protect\citeauthoryear{{Naik}, {Puchwein}, {Davis} \&
  {Arnold}}{{Naik} et~al.}{2018}]{Naik_etal_2018}
{Naik} A.~P.,  {Puchwein} E.,  {Davis} A.-C.,    {Arnold} C.,  2018, ArXiv
  e-prints 1805.12221

\bibitem[\protect\citeauthoryear{Novikov}{Novikov}{2016}]{novikov16b}
Novikov E.~A.,  2016, Electron. J. Theor. Phys., 13, 79

\bibitem[\protect\citeauthoryear{Oyaizu, Lima \& Hu}{Oyaizu
  et~al.}{2008}]{Oyaizu_etal_2008}
Oyaizu H.,  Lima M.,    Hu W.,  2008, Phys. Rev., D78, 123524

\bibitem[\protect\citeauthoryear{{Peel}, {Pettorino}, {Giocoli}, {Starck} \&
  {Baldi}}{{Peel} et~al.}{2018}]{peel18}
{Peel} A.,  {Pettorino} V.,  {Giocoli} C.,  {Starck} J.-L.,    {Baldi} M.,
  2018, eprints arXiv: 1805.05146

\bibitem[\protect\citeauthoryear{{Petri}, {Haiman} \& {May}}{{Petri}
  et~al.}{2016}]{petri16}
{Petri} A.,  {Haiman} Z.,    {May} M.,  2016, \prd, 93, 063524

\bibitem[\protect\citeauthoryear{{Petri}, {Haiman} \& {May}}{{Petri}
  et~al.}{2017}]{petri17}
{Petri} A.,  {Haiman} Z.,    {May} M.,  2017, \prd, 95, 123503

\bibitem[\protect\citeauthoryear{{Planck Collaboration}}{{Planck
  Collaboration}}{2016}]{planck16a}
{Planck Collaboration} 2016, Astron. Astrophys., 594, A13

\bibitem[\protect\citeauthoryear{{Planck Collaboration}, {Adam}, {Ade},
  {Aghanim}, {Akrami}, {Alves}, {Arnaud}, {Arroja}, {Aumont}, {Baccigalupi} \&
  et al.}{{Planck Collaboration} et~al.}{2015}]{planck1_15}
{Planck Collaboration} {Adam} R.,  {Ade} P.~A.~R.,  {Aghanim} N.,  {Akrami} Y.,
   {Alves} M.~I.~R.,  {Arnaud} M.,  {Arroja} F.,  {Aumont} J.,  {Baccigalupi}
  C.,    et al. 2015, ArXiv e-prints: 1502.01582

\bibitem[\protect\citeauthoryear{{Planck Collaboration}, {Ade}, {Aghanim},
  {Alves}, {Armitage-Caplan}, {Arnaud}, {Ashdown}, {Atrio-Barandela}, {Aumont},
  {Aussel} \& et al.}{{Planck Collaboration} et~al.}{2014}]{planck1_14}
{Planck Collaboration} {Ade} P.~A.~R.,  {Aghanim} N.,  {Alves} M.~I.~R.,
  {Armitage-Caplan} C.,  {Arnaud} M.,  {Ashdown} M.,  {Atrio-Barandela} F.,
  {Aumont} J.,  {Aussel} H.,    et al. 2014, \aap, 571, A1

\bibitem[\protect\citeauthoryear{{Planck Collaboration}, {Ade}, {Aghanim},
  {Armitage-Caplan}, {Arnaud}, {Ashdown}, {Atrio-Barandela}, {Aumont},
  {Baccigalupi}, {Banday} \& et al.}{{Planck Collaboration}
  et~al.}{2014}]{planckxx}
{Planck Collaboration} {Ade} P.~A.~R.,  {Aghanim} N.,  {Armitage-Caplan} C.,
  {Arnaud} M.,  {Ashdown} M.,  {Atrio-Barandela} F.,  {Aumont} J.,
  {Baccigalupi} C.,  {Banday} A.~J.,    et al. 2014, \aap, 571, A20

\bibitem[\protect\citeauthoryear{{Planck Collaboration}, {Ade}, {Aghanim},
  {Arnaud}, {Ashdown}, {Atrio-Barandela}, {Aumont}, {Baccigalupi}, {Balbi},
  {Banday} \& et al.}{{Planck Collaboration} et~al.}{2013}]{planckxi}
{Planck Collaboration} {Ade} P.~A.~R.,  {Aghanim} N.,  {Arnaud} M.,  {Ashdown}
  M.,  {Atrio-Barandela} F.,  {Aumont} J.,  {Baccigalupi} C.,  {Balbi} A.,
  {Banday} A.~J.,    et al. 2013, \aap, 557, A52

\bibitem[\protect\citeauthoryear{{Planck Collaboration}, {Ade}, {Aghanim},
  {Arnaud}, {Ashdown}, {Aumont}, {Baccigalupi}, {Baker}, {Balbi}, {Banday} \&
  et al.}{{Planck Collaboration} et~al.}{2011}]{planck1_11}
{Planck Collaboration} {Ade} P.~A.~R.,  {Aghanim} N.,  {Arnaud} M.,  {Ashdown}
  M.,  {Aumont} J.,  {Baccigalupi} C.,  {Baker} M.,  {Balbi} A.,  {Banday}
  A.~J.,    et al. 2011, \aap, 536, A1

\bibitem[\protect\citeauthoryear{{Planck Collaboration}, {Ade}, {Aghanim},
  {Arnaud}, {Ashdown}, {Aumont}, {Baccigalupi}, {Banday}, {Barreiro},
  {Bartlett} \& et al.}{{Planck Collaboration}
  et~al.}{2016b}]{Planck_2015_XIII}
{Planck Collaboration} {Ade} P.~A.~R.,  {Aghanim} N.,  {Arnaud} M.,  {Ashdown}
  M.,  {Aumont} J.,  {Baccigalupi} C.,  {Banday} A.~J.,  {Barreiro} R.~B.,
  {Bartlett} J.~G.,    et al. 2016b, \aap, 594, A13

\bibitem[\protect\citeauthoryear{{Planck Collaboration}, {Ade}, {Aghanim},
  {Arnaud}, {Ashdown}, {Aumont}, {Baccigalupi}, {Banday}, {Barreiro},
  {Bartlett} \& et al.}{{Planck Collaboration} et~al.}{2016a}]{planckxxiv}
{Planck Collaboration} {Ade} P.~A.~R.,  {Aghanim} N.,  {Arnaud} M.,  {Ashdown}
  M.,  {Aumont} J.,  {Baccigalupi} C.,  {Banday} A.~J.,  {Barreiro} R.~B.,
  {Bartlett} J.~G.,    et al. 2016a, \aap, 594, A24

\bibitem[\protect\citeauthoryear{{Poulin}, {Boddy}, {Bird} \&
  {Kamionkowski}}{{Poulin} et~al.}{2018}]{poulin18}
{Poulin} V.,  {Boddy} K.~K.,  {Bird} S.,    {Kamionkowski} M.,  2018, ArXiv
  e-prints: 1803.02474

\bibitem[\protect\citeauthoryear{{Puchwein}, {Baldi} \& {Springel}}{{Puchwein}
  et~al.}{2013}]{Puchwein_Baldi_Springel_2013}
{Puchwein} E.,  {Baldi} M.,    {Springel} V.,  2013, \mnras, 436, 348

\bibitem[\protect\citeauthoryear{{Refregier}, {Massey}, {Rhodes}, {Ellis},
  {Albert}, {Bacon}, {Bernstein}, {McKay} \& {Perlmutter}}{{Refregier}
  et~al.}{2004}]{refregier04}
{Refregier} A.,  {Massey} R.,  {Rhodes} J.,  {Ellis} R.,  {Albert} J.,  {Bacon}
  D.,  {Bernstein} G.,  {McKay} T.,    {Perlmutter} S.,  2004, \aj, 127, 3102

\bibitem[\protect\citeauthoryear{{Roncarelli}, {Baldi} \&
  {Villaescusa-Navarro}}{{Roncarelli}
  et~al.}{2018}]{Roncarelli_Baldi_Villaescusa-Navarro_2018}
{Roncarelli} M.,  {Baldi} M.,    {Villaescusa-Navarro} F.,  2018, ArXiv
  e-prints

\bibitem[\protect\citeauthoryear{{Roncarelli}, {Moscardini}, {Borgani} \&
  {Dolag}}{{Roncarelli} et~al.}{2007}]{roncarelli07}
{Roncarelli} M.,  {Moscardini} L.,  {Borgani} S.,    {Dolag} K.,  2007, \mnras,
  378, 1259

\bibitem[\protect\citeauthoryear{{Sartoris}, {Biviano}, {Fedeli}, {Bartlett},
  {Borgani}, {Costanzi}, {Giocoli}, {Moscardini}, {Weller}, {Ascaso},
  {Bardelli}, {Maurogordato} \& {Viana}}{{Sartoris} et~al.}{2016}]{sartoris16}
{Sartoris} B.,  {Biviano} A.,  {Fedeli} C.,  {Bartlett} J.~G.,  {Borgani} S.,
  {Costanzi} M.,  {Giocoli} C.,  {Moscardini} L.,  {Weller} J.,  {Ascaso} B.,
  {Bardelli} S.,  {Maurogordato} S.,    {Viana} P.~T.~P.,  2016, \mnras, 459,
  1764

\bibitem[\protect\citeauthoryear{{Sch{\"a}fer}, {Heisenberg}, {Kalovidouris} \&
  {Bacon}}{{Sch{\"a}fer} et~al.}{2012}]{schaefer12}
{Sch{\"a}fer} B.~M.,  {Heisenberg} L.,  {Kalovidouris} A.~F.,    {Bacon} D.~J.,
   2012, \mnras, 420, 455

\bibitem[\protect\citeauthoryear{{Schmidt}}{{Schmidt}}{2008}]{schmidt08}
{Schmidt} F.,  2008, \prd, 78, 043002

\bibitem[\protect\citeauthoryear{Schmidt, Lima, Oyaizu \& Hu}{Schmidt
  et~al.}{2009}]{Schmidt_etal_2009}
Schmidt F.,  Lima M.~V.,  Oyaizu H.,    Hu W.,  2009, Phys. Rev., D79, 083518

\bibitem[\protect\citeauthoryear{{Seitz} \& {Schneider}}{{Seitz} \&
  {Schneider}}{1997}]{seitz97}
{Seitz} C.,  {Schneider} P.,  1997, \aap, 318, 687

\bibitem[\protect\citeauthoryear{{Shan}, {Liu}, {Hildebrandt}, {Pan},
  {Martinet}, {Fan}, {Schneider}, {Asgari} \& {et al.}}{{Shan}
  et~al.}{2018}]{shan17}
{Shan} H.,  {Liu} X.,  {Hildebrandt} H.,  {Pan} C.,  {Martinet} N.,  {Fan} Z.,
  {Schneider} P.,  {Asgari} M.,    {et al.} 2018, \mnras, 474, 1116

\bibitem[\protect\citeauthoryear{{Shirasaki}, {Nishimichi}, {Li} \&
  {Higuchi}}{{Shirasaki} et~al.}{2017}]{shirasaki17}
{Shirasaki} M.,  {Nishimichi} T.,  {Li} B.,    {Higuchi} Y.,  2017, \mnras,
  466, 2402

\bibitem[\protect\citeauthoryear{Sotiriou \& Faraoni}{Sotiriou \&
  Faraoni}{2010}]{Sotiriou_Faraoni_2010}
Sotiriou T.~P.,  Faraoni V.,  2010, Rev.Mod.Phys., 82, 451

\bibitem[\protect\citeauthoryear{Springel}{Springel}{2005}]{gadget-2}
Springel V.,  2005, Mon. Not. Roy. Astron. Soc., 364, 1105

\bibitem[\protect\citeauthoryear{{Springel}, {White}, {Tormen} \&
  {Kauffmann}}{{Springel} et~al.}{2001}]{Springel_etal_2001}
{Springel} V.,  {White} S.~D.~M.,  {Tormen} G.,    {Kauffmann} G.,  2001,
  \mnras, 328, 726

\bibitem[\protect\citeauthoryear{Starobinsky}{Starobinsky}{1980}]{Starobinsky_1980}
Starobinsky A.~A.,  1980, Phys. Lett., B91, 99

\bibitem[\protect\citeauthoryear{{Takahashi}, {Sato}, {Nishimichi}, {Taruya} \&
  {Oguri}}{{Takahashi} et~al.}{2012}]{takahashi12}
{Takahashi} R.,  {Sato} M.,  {Nishimichi} T.,  {Taruya} A.,    {Oguri} M.,
  2012, \apj, 761, 152

\bibitem[\protect\citeauthoryear{Valageas}{Valageas}{2000}]{valageas00}
Valageas P.,  2000, A\&A, 356, 771

\bibitem[\protect\citeauthoryear{{Vicinanza}, {Cardone}, {Maoli}, {Scaramella}
  \& {Er}}{{Vicinanza} et~al.}{2018}]{vicinanza18}
{Vicinanza} M.,  {Cardone} V.~F.,  {Maoli} R.,  {Scaramella} R.,    {Er} X.,
  2018, \prd, 97, 023519

\bibitem[\protect\citeauthoryear{Viel, Haehnelt \& Springel}{Viel
  et~al.}{2010}]{Viel_Haehnelt_Springel_2010}
Viel M.,  Haehnelt M.~G.,    Springel V.,  2010, JCAP, 1006, 015

\bibitem[\protect\citeauthoryear{{Villaescusa-Navarro}, {Banerjee}, {Dalal},
  {Castorina}, {Scoccimarro}, {Angulo} \& {Spergel}}{{Villaescusa-Navarro}
  et~al.}{2017}]{Villaescusa-Navarro_etal_2018}
{Villaescusa-Navarro} F.,  {Banerjee} A.,  {Dalal} N.,  {Castorina} E.,
  {Scoccimarro} R.,  {Angulo} R.,    {Spergel} D.~N.,  2017, ArXiv e-prints

\bibitem[\protect\citeauthoryear{{Wagner}, {Verde} \& {Jimenez}}{{Wagner}
  et~al.}{2012}]{Wagner_Verde_Jimenez_2012}
{Wagner} C.,  {Verde} L.,    {Jimenez} R.,  2012, \apjl, 752, L31

\bibitem[\protect\citeauthoryear{Will}{Will}{2005}]{Will_2005}
Will C.~M.,  2005, Living Rev.Rel., 9, 3

\bibitem[\protect\citeauthoryear{Winther et~al.,}{Winther
  et~al.}{2015}]{Winther_etal_2015}
Winther H.~A.,  et~al., 2015, Mon. Not. Roy. Astron. Soc., 454, 4208

\bibitem[\protect\citeauthoryear{{Wright}, {Winther} \& {Koyama}}{{Wright}
  et~al.}{2017}]{Wright_Winther_Koyama_2017}
{Wright} B.~S.,  {Winther} H.~A.,    {Koyama} K.,  2017, \jcap, 10, 054

\bibitem[\protect\citeauthoryear{{Zennaro}, {Bel}, {Villaescusa-Navarro},
  {Carbone}, {Sefusatti} \& {Guzzo}}{{Zennaro}
  et~al.}{2017}]{Zennaro_etal_2017}
{Zennaro} M.,  {Bel} J.,  {Villaescusa-Navarro} F.,  {Carbone} C.,  {Sefusatti}
  E.,    {Guzzo} L.,  2017, \mnras, 466, 3244

\bibitem[\protect\citeauthoryear{Zhao, Li \& Koyama}{Zhao
  et~al.}{2011}]{Zhao_Li_Koyama_2011a}
Zhao G.-B.,  Li B.,    Koyama K.,  2011, Phys.Rev., D83, 044007

\bibitem[\protect\citeauthoryear{{Zhao}, {Pogosian}, {Silvestri} \&
  {Zylberberg}}{{Zhao} et~al.}{2009}]{zhao09a}
{Zhao} G.-B.,  {Pogosian} L.,  {Silvestri} A.,    {Zylberberg} J.,  2009, \prd,
  79, 083513

\bibitem[\protect\citeauthoryear{{Zorrilla Matilla}, {Haiman}, {Hsu}, {Gupta}
  \& {Petri}}{{Zorrilla Matilla} et~al.}{2016}]{matilla16}
{Zorrilla Matilla} J.~M.,  {Haiman} Z.,  {Hsu} D.,  {Gupta} A.,    {Petri} A.,
  2016, \prd, 94, 083506

\end{thebibliography}
\label{lastpage}
\end{document}